\documentclass{article}

\makeatletter

\relax

\usepackage[verbose=true,letterpaper]{geometry}
\AtBeginDocument{
  \newgeometry{
    textheight=9in,
    textwidth=6.5in,
    top=1in,
    headheight=14pt,
    headsep=25pt,
    footskip=30pt
  }
}

\usepackage{fancyhdr}
\fancyheadoffset{0pt}
\def\keywordname{{\bfseries \emph Keywords}}%
\def\keywords#1{\par\addvspace\medskipamount{\rightskip=0pt plus1cm
\def\and{\ifhmode\unskip\nobreak\fi\ $\cdot$
}\noindent\keywordname\enspace\ignorespaces#1\par}}

\renewcommand{\normalsize}{%
  \@setfontsize\normalsize\@xpt\@xipt
  \abovedisplayskip      7\p@ \@plus 2\p@ \@minus 5\p@
  \abovedisplayshortskip \z@ \@plus 3\p@
  \belowdisplayskip      \abovedisplayskip
  \belowdisplayshortskip 4\p@ \@plus 3\p@ \@minus 3\p@
}
\normalsize
\renewcommand{\small}{%
  \@setfontsize\small\@ixpt\@xpt
  \abovedisplayskip      6\p@ \@plus 1.5\p@ \@minus 4\p@
  \abovedisplayshortskip \z@  \@plus 2\p@
  \belowdisplayskip      \abovedisplayskip
  \belowdisplayshortskip 3\p@ \@plus 2\p@   \@minus 2\p@
}
\renewcommand{\footnotesize}{\@setfontsize\footnotesize\@ixpt\@xpt}
\renewcommand{\scriptsize}{\@setfontsize\scriptsize\@viipt\@viiipt}
\renewcommand{\tiny}{\@setfontsize\tiny\@vipt\@viipt}
\renewcommand{\large}{\@setfontsize\large\@xiipt{14}}
\renewcommand{\Large}{\@setfontsize\Large\@xivpt{16}}
\renewcommand{\LARGE}{\@setfontsize\LARGE\@xviipt{20}}
\renewcommand{\huge}{\@setfontsize\huge\@xxpt{23}}
\renewcommand{\Huge}{\@setfontsize\Huge\@xxvpt{28}}

\providecommand{\section}{}
\renewcommand{\section}{%
  \@startsection{section}{1}{\z@}%
                {-2.0ex \@plus -0.5ex \@minus -0.2ex}%
                { 1.5ex \@plus  0.3ex \@minus  0.2ex}%
                {\large\bf\raggedright}%
}
\providecommand{\subsection}{}
\renewcommand{\subsection}{%
  \@startsection{subsection}{2}{\z@}%
                {-1.8ex \@plus -0.5ex \@minus -0.2ex}%
                { 0.8ex \@plus  0.2ex}%
                {\normalsize\bf\raggedright}%
}
\providecommand{\subsubsection}{}
\renewcommand{\subsubsection}{%
  \@startsection{subsubsection}{3}{\z@}%
                {-1.5ex \@plus -0.5ex \@minus -0.2ex}%
                { 0.5ex \@plus  0.2ex}%
                {\normalsize\bf\raggedright}%
}
\providecommand{\paragraph}{}
\renewcommand{\paragraph}{%
  \@startsection{paragraph}{4}{\z@}%
                {1.5ex \@plus 0.5ex \@minus 0.2ex}%
                {-1em}%
                {\normalsize\bf}%
}
\providecommand{\subparagraph}{}
\renewcommand{\subparagraph}{%
  \@startsection{subparagraph}{5}{\z@}%
                {1.5ex \@plus 0.5ex \@minus 0.2ex}%
                {-1em}%
                {\normalsize\bf}%
}

\newlength{\@abovecaptionskip}
\newlength{\@belowcaptionskip}

\renewenvironment{table}
  {\setlength{\abovecaptionskip}{\@belowcaptionskip}%
   \setlength{\belowcaptionskip}{\@abovecaptionskip}%
   \@float{table}}
  {\end@float}

\renewcommand{\footnoterule}{\kern-3\p@ \hrule width 12pc \kern 2.6\p@}
\def\@listi  {\leftmargin\leftmargini}
\def\@listii {\leftmargin\leftmarginii
              \labelwidth\leftmarginii
              \advance\labelwidth-\labelsep
              \topsep  2\p@ \@plus 1\p@    \@minus 0.5\p@
              \parsep  1\p@ \@plus 0.5\p@ \@minus 0.5\p@
              \itemsep \parsep}
\def\@listiii{\leftmargin\leftmarginiii
              \labelwidth\leftmarginiii
              \advance\labelwidth-\labelsep
              \topsep    1\p@ \@plus 0.5\p@ \@minus 0.5\p@
              \parsep    \z@
              \partopsep 0.5\p@ \@plus 0\p@ \@minus 0.5\p@
              \itemsep \topsep}
\def\@listiv {\leftmargin\leftmarginiv
              \labelwidth\leftmarginiv
              \advance\labelwidth-\labelsep}
\def\@listv  {\leftmargin\leftmarginv
              \labelwidth\leftmarginv
              \advance\labelwidth-\labelsep}
\def\@listvi {\leftmargin\leftmarginvi
              \labelwidth\leftmarginvi
              \advance\labelwidth-\labelsep}

\providecommand{\maketitle}{}
\renewcommand{\maketitle}{%
  \par
  \begingroup
    \renewcommand{\thefootnote}{\fnsymbol{footnote}}
    \renewcommand{\@makefnmark}{\hbox to \z@{$^{\@thefnmark}$\hss}}
    \long\def\@makefntext##1{%
      \parindent 1em\noindent
      \hbox to 1.8em{\hss $\m@th ^{\@thefnmark}$}##1
    }
    \thispagestyle{empty}
    \@maketitle
    \@thanks
  \endgroup
  \let\maketitle\relax
  \let\thanks\relax
}

\newcommand{\@toptitlebar}{
  \hrule height 2\p@
  \vskip 0.25in
  \vskip -\parskip%
}
\newcommand{\@bottomtitlebar}{
  \vskip 0.29in
  \vskip -\parskip
  \hrule height 2\p@
  \vskip 0.09in%
}

\providecommand{\@maketitle}{}
\renewcommand{\@maketitle}{%
  \vbox{%
    \hsize\textwidth
    \linewidth\hsize
    \vskip 0.1in
    \@toptitlebar
    \centering
    {\LARGE\sc \@title\par}
    \@bottomtitlebar
    \vskip 0.1in
    \def\And{%
      \end{tabular}\hfil\linebreak[0]\hfil%
      \begin{tabular}[t]{c}\bf\rule{\z@}{24\p@}\ignorespaces%
    }
    \def\AND{%
      \end{tabular}\hfil\linebreak[4]\hfil%
      \begin{tabular}[t]{c}\bf\rule{\z@}{24\p@}\ignorespaces%
    }
    \begin{tabular}[t]{c}\bf\rule{\z@}{24\p@}\@author\end{tabular}%
  \vskip 0.4in \@minus 0.1in \center{ }   \vskip 0.2in 
  }
}

\newcommand{\ftype@noticebox}{8}
\newcommand{\@notice}{%
  \enlargethispage{2\baselineskip}%
  \begin{noticebox}[b]
    \footnotesize\@noticestring
  \end{noticebox}
}

\renewenvironment{abstract}
{
  \centerline
  {\large \bfseries \scshape Abstract}
  \begin{quote}
}
{
  \end{quote}
}

\makeatother

\usepackage[utf8]{inputenc} 
\usepackage[T1]{fontenc}    
\usepackage{hyperref}       
\usepackage{url}            
\usepackage{booktabs}       
\usepackage{amsfonts}       
\usepackage{nicefrac}       
\usepackage{microtype}      
\usepackage{lipsum}
\usepackage{fancyhdr}       
\usepackage{graphicx}       
\graphicspath{{media/}}     
\usepackage{array}
\usepackage{multirow}
\usepackage{tabularx}
\usepackage{tikz}
\usepackage{adjustbox}
\usepackage{float}
\usepackage{placeins}

\usepackage{amsmath}
\usepackage{subcaption}

\begin{document}

\title{Future Pathways for eVTOLs: \\[1.5ex] \LARGE{A Design Optimization Perspective}

}

\begin{center}
  \author{
    Johannes Janning, Sophie F. Armanini, Urban Fasel \\
    Department of Aeronautics, Imperial College London, United Kingdom \\
    \texttt{\{johannes.janning23, s.armanini, u.fasel, \}@imperial.ac.uk}
  }
\end{center}

\vspace{-2em}

\maketitle
\begin{abstract}
The rapid development of advanced urban air mobility, particularly electric vertical take-off and landing (eVTOL) aircraft, requires interdisciplinary approaches involving the future urban air mobility ecosystem. Operational cost efficiency, regulatory aspects, sustainability, and environmental compatibility should be incorporated directly into the conceptual design of aircraft and across operational and regulatory strategies. In this work, we apply a multidisciplinary design optimization framework for the conceptual design of eVTOL aircraft. The framework optimizes conventional design elements of eVTOL aircraft over a generic mission and integrates operational cost and an energy and battery-based emissions model to capture sustainability incentives of the designed system. We introduce a novel metric, the cross-transportation Figure of Merit (FoM), to compare the optimized eVTOL system with various competing road, rail, and air transportation modes in terms of sustainability, cost, and travel time. The framework incorporates regulatory, technical, and operational constraints to evaluate four distinct stakeholder-centric configurations: profit-maximizing, cost-minimizing, environmental impact-minimizing, and FoM-optimized designs. Our analysis shows that prioritizing environmental impact can reduce combined energy and battery-related emissions by over 50\% compared to profit-maximizing designs. However, this comes at the cost of longer trip times and lower annual profits, driven by reduced vehicle utilization. In contrast, profit-maximizing designs achieve strong economic performance through rapid battery charging and high flight frequency, but result in significantly accelerated battery degradation and can reduce battery life by 85\%. These results highlight the importance of integrating multiple operational disciplines into the design process, while underscoring the differing priorities of operators, regulators, and society.

\paragraph{Highlights:}
\begin{itemize}
    \item Integrated multidisciplinary design optimization framework for eVTOLs in UAM ecosystem.
    \item Critical design and cost trade-offs identified in eVTOL operations.
    \item Figure of Merit proposed as new benchmark for cross-modal comparison.
    \item Actionable strategies provided for urban air mobility stakeholders.
    \item Scalable eVTOL design, operations, and economic models provided.
\end{itemize}
\end{abstract}


\newpage

\section*{List of Abbreviations}
\addcontentsline{toc}{section}{List of Abbreviations}

\begin{tabularx}{\textwidth}{@{}lX@{}}
\textbf{Abbreviation} & \textbf{Description} \\
& \\
AGL & Above Ground Level \\
ASK & Available Seat Kilometers \\
ATA & Air Transport Association \\
COC & Cash Operating Cost \\
COO & Cost of Ownership \\
DEP & Distributed Electric Propulsion \\
DOC & Direct Operating Cost \\
DoD & Depth of Discharge \\
EASA & European Union Aviation Safety Agency \\
EOL & End-of-Life \\
eVTOL & electric Vertical Take-Off and Landing \\
EV & Electric Vehicle \\
FAA & Federal Aviation Administration \\
FEM & Finite Element Method \\
FoM & Figure of Merit \\
GWP & Global Warming Potential \\
IOC & Indirect Operating Cost \\
MDO & Multidisciplinary Design Optimization \\
MTOM & Maximum Take-Off Mass \\
NASA & National Aeronautics and Space Administration \\
NCM & Nickel-Cobalt-Manganese \\
PAX & Passengers \\
SFAR & Special Federal Aviation Regulation \\
SLSQP & Sequential Least Squares Programming \\
SOC & State-of-Charge \\
SPL & Sound Pressure Level \\
TOC & Total Operating Cost \\
UAM & Urban Air Mobility \\
\end{tabularx}

\newpage

\section{Introduction}

Urban air mobility (UAM) envisions the use of electric vertical take-off and landing aircraft (eVTOLs) to transport passengers and cargo within urban and regional networks \cite{hollistic}. These aircraft are intended to provide fast, efficient, and sustainable mobility over short to medium distances. More broadly, eVTOLs will serve as the core of an integrated ecosystem comprising ground and airborne infrastructure, operators, and regulators.
Like helicopters, eVTOLs have vertical flight capabilities that enable operation in dense urban environments. However, they are powered by quieter and cleaner distributed electric propulsion (DEP) systems \cite{Schaefer2018}. Many configurations also enable wing-borne cruise flight, combining aerodynamic efficiency with operational versatility. Recent advances in electric systems and battery technology have made eVTOL concepts technically viable. Nevertheless, most designs remain at the prototype stage, and their full integration into the aviation ecosystem has yet to be realized. Their success will depend on meeting criteria such as low energy consumption, reduced emissions, and cost-effective operations, all of which are critical for achieving broad public acceptance \cite{EASA2021social}. 
Addressing these requirements requires design frameworks that simultaneously consider aircraft sizing, operational economics, and environmental impact. However, most existing optimization studies focus on isolated performance metrics \cite{Martins2023, Sarojini2023, Ruh2024, Keijzer2024}, often overlooking the trade-offs between operational objectives such as profit maximization and environmental sustainability. Consequently, there is a need for a holistic framework capable of benchmarking eVTOL designs against alternative transportation systems. 

In this work, we address this need by developing a multidisciplinary design optimization (MDO) framework for the conceptual design of eVTOL aircraft. The framework integrates vehicle sizing, emission-based sustainability, and operational economics, enabling the analysis and optimization of technological, environmental, regulatory, and economic interactions in the eVTOL ecosystem. It supports the evaluation of different scenarios, such as maximizing profit, minimizing cost, or reducing global warming potential, to derive practical insights for urban air mobility stakeholders.
We demonstrate the approach using a 70-km sizing mission, in which different regulatory and industry-specific scenarios are examined. The analysis accounts for both operational and design constraints, including charging rates, rotor design, and vertiport requirements. To enable comparison with alternative transportation modes, we introduce a new metric, the cross-transportation Figure of Merit (FoM), which balances travel time, environmental impact, and cost. This facilitates comparison of eVTOLs to ground-, rail-, and air-based competitors. This metric provides guidance for the development and deployment of eVTOL systems that are both sustainable and economically viable, while considering stakeholder-driven design preferences. Although the present analysis focuses on a representative 70-km mission, the framework is flexible and scalable, and can be applied to a wide range of mission distances.

The remainder of this paper is organized as follows: Section \ref{ch:literature} reviews related literature. Section \ref{ch:met} details the multidisciplinary design optimization framework and the underlying models, with Subsection \ref{ch:fom} introducing the cross-transportation Figure of Merit. Section \ref{ch:results} presents the results of the optimization, followed by a discussion of the findings in Section \ref{ch:discussion}. Finally, Section \ref{ch:conclusion} provides concluding remarks and outlook for future work.

\begin{figure}[ht]
    \centering
    \includegraphics[width=\textwidth]{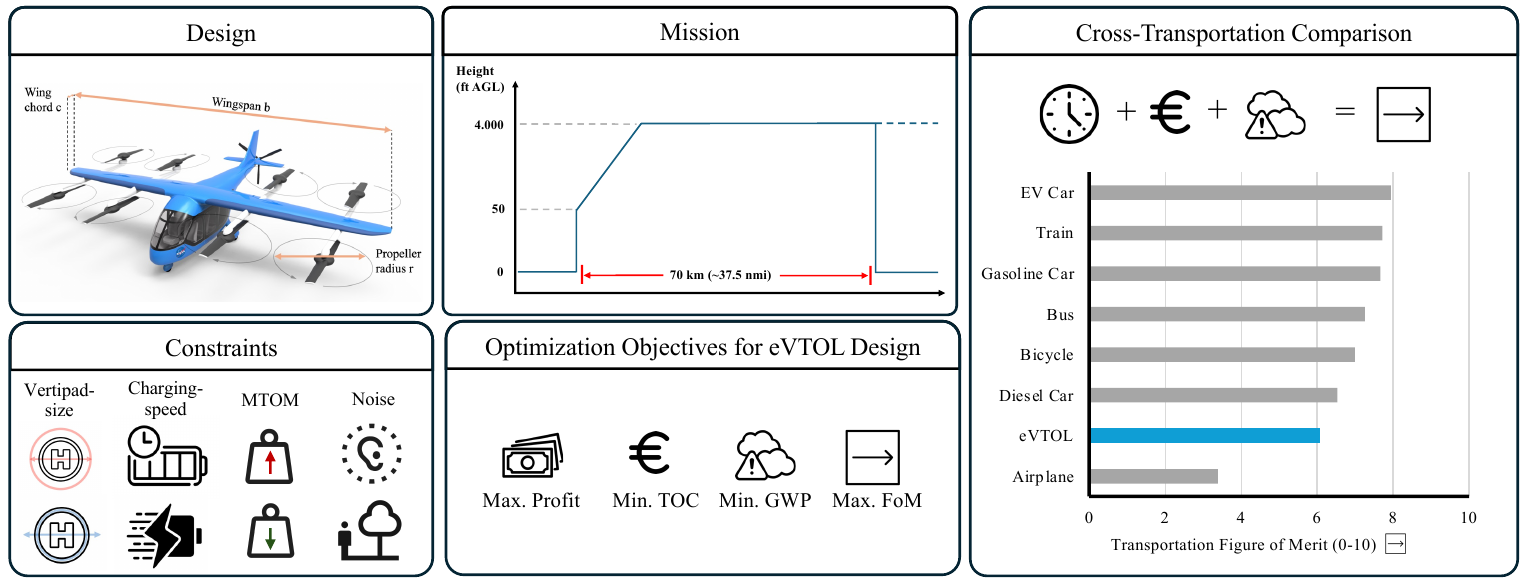} 
    \caption{Framework for the conceptual eVTOL design optimization, illustrating the key geometric parameters, the mission profile, and the operational constraints. It outlines the four primary optimization objectives: maximum profit, minimum total operating cost (TOC), minimum Global Warming Potential (GWP), and maximum stakeholder-centric Figure of Merit (FoM), which are compared against conventional transportation modes (illustration adapted from \cite{nasa_uam_refs, AIP_SkyVector}).}
    \label{fig:01}
\end{figure}

\section{Related Work}
\label{ch:literature}

Developing a comprehensive MDO framework for eVTOLs is crucial for addressing the complex trade-offs in UAM. Existing research primarily focuses on the optimization of aerodynamic performance, structural properties, and mission efficiency. Table \ref{tab:sliterature} summarizes relevant eVTOL design and mission optimization literature, identified as related work. Several recent studies have achieved reductions in vehicle weight and improvements in energy efficiency, which are both crucial for operational feasibility~\cite{Martins2023, Sarojini2023}. Despite extensive MDO research \cite{Martins2023, martins2021engineering, Sarojini2023, Ruh2024, Keijzer2024}, however, critical aspects such as economic viability and environmental impact have been less frequently addressed. Other studies suggest potential enhancements in mission efficiency and cost reduction but also expose notable gaps in addressing operational factors, particularly in sustainability, operations modeling, and vehicle utilization \cite{Ha2019, Chinthoju2024}. Areas such as carbon footprint and battery lifecycle for eVTOL operations are rarely considered \cite{brown2018vehicle, hollistic}. Clearly, there is a need for a more integrated approach that incorporates profitability and other objectives of potential future eVTOL operators into early-stage design optimization.

\begin{table}[h]
    \centering
    \footnotesize
    \begin{tabular}{lll}
        \hline\hline
        \textbf{Authors} & \textbf{Year} & \textbf{Objective}  \\
        \hline\hline
        Perez et al. \cite{Perez2025} & 2025 & Assess economic, energy, and environmental feasibility of eVTOL commuting. \\
        Ruh et al. \cite{Ruh2024} & 2024 & Integrate physics-based models in MDO to optimize design. \\
        Chinthoju et al. \cite{Chinthoju2024} & 2024 & Cost optimization by varying mass, rotor radius, wing span, cruise speed. \\
        Lio et al. \cite{Liu2024}&2024 & Techno-economic feasibility analysis of eVTOL air taxis. \\
        Khavarian et al. \cite{Khavarian2024LCA} & 2024 & LCA of eVTOL costs and emissions in Austin network. \\
        Fioriti et al. \cite{Fioriti2024ICAS} & 2024 & Parametric cost and LCA models for eVTOL fleets. \\
        Kaneko et al. \cite{Martins2023} & 2023 & Optimize eVTOL aerodynamics and energy efficiency for weight reduction. \\
        Sarojini et al. \cite{Sarojini2023} & 2023 & Optimize lift+cruise \cite{patterson2018proposed} weight using a geometry-centric approach. \\
        Kiesewetter et al. \cite{hollistic} & 2023 & Holistic UAM ecosystem state of research review \& evaluation. \\
        Straubinger et al. \cite{Straubinger2023UAM} & 2023 & Critique of UAM sustainability and equity impacts. \\
        Hagag et al. \cite{Hagag2023} & 2023 & Compare eVTOL time savings and CO$_2$ vs. cars, trains, and helicopters (Paris). \\
        Straubinger et al. \cite{Straubinger2022GoingElectric} & 2022 & Urban equilibrium model of UAM, CO$_2$ and welfare effects. \\
        Donateo et al. \cite{Donateo2022} & 2022 & Compare hybrid/all-electric eVTOLs with electric road vehicles in taxi services. \\
        Yang et al. \cite{yang2021challenges} & 2021 & Impact of charging strategies on battery sizing and utilization. \\
        Mihara et al. \cite{Mihara2021} & 2021 & Cost analysis for diverse eVTOL configurations. \\
        Ha et al. \cite{Ha2019} & 2019 & Optimize mission efficiency and profitability in eVTOL designs. \\
        Kasliwal et al. \cite{kasliwal2019role} & 2019 & Physics-based analysis of VTOLs emissions vs. ground-based cars. \\
        Schäfer et al. \cite{Schaefer2018} & 2019 & Energy, economic and environmental implications of all-electric aircraft. \\ 
        André et al. \cite{Andre2019} & 2019 & Comprehensive eVTOL life-cycle assessment and design implications. \\ 
        Brown et al. \cite{brown2018vehicle} & 2018 & Design \& economic optimization using geometric programming. \\
        Silva et al. \cite{silva2018} & 2018 & Examination of lift+cruise vehicle for research \& industry guidance. \\ 
        Patterson et al. \cite{patterson2018proposed} & 2018 & Guideline to study UAM missions \& exploration of mission requirements. \\
        \hline\hline
    \end{tabular}
    \vspace{6pt}
    \caption{Related eVTOL design and mission optimization literature.}
    \label{tab:sliterature}
\end{table}

Table \ref{tab:sliterature} summarizes related eVTOL design and mission optimization work. Brown et al. \cite{brown2018vehicle} employed geometric programming to optimize vehicle components and mission profiles, including noise calculations and a general cost model. However, the study provided limited detail on cost breakdowns for infrastructure, regulatory compliance, and long-term maintenance. Additionally, it did not fully address environmental costs related to energy production and battery degradation, which are essential for a comprehensive evaluation of the economic and environmental sustainability of UAM. Recent reviews have highlighted significant gaps in integrating economic models for evaluating cost-effectiveness, lifecycle costs, and environmental impacts in eVTOL design~\cite{hollistic}. 

The eVTOL analysis framework developed in~\cite{silva2018} and \cite{patterson2018proposed} provides insightful guidelines for eVTOL aircraft architecture and mission design that can be adapted to short- and mid-range lift and cruise eVTOL missions. Key insights on battery behavior and its implications for operational charging strategies can be drawn from~\cite{yang2021challenges} and \cite{kasliwal2019role}, and critical perspectives on the interplay between economics, technology, and sustainability in fully electric aircraft are offered in~\cite{Schaefer2018}, highlighting their interlinked nature, with \cite{Andre2019} and \cite{Arshad2022} providing essential considerations for the environmental impacts of battery usage. 
A framework for standardized cost model structuring is provided in \cite{Faulkner1973}, and valuable input on revenue modeling is provided in \cite{Mihara2021, Liu2024}. Finally, economic key requirements are defined for real-world operations in \cite{uberelevate}, useful for validation in eVTOL MDO. 

Several studies with fixed eVTOL designs examine operations, economics, and broader societal impacts. At the network level, \cite{Khavarian2024LCA} assesses deployment in Austin, Texas, considering vertiport service areas and demand thresholds, finding that small (4-seat) vehicles can reduce emissions relative to larger types but remain cost-intensive versus ground-based electric vehicles (EV). Early-stage fleet cost and life cycle assessment (LCA) models by \cite{Fioriti2024ICAS} highlight maintenance, especially battery replacement, as a dominant operating cost and electricity generation as the main driver of life-cycle emissions, underscoring dependence on grid mix. From a sustainability and equity perspective, \cite{Straubinger2023UAM} argues that UAM is not inherently green, as its system-level impact may be limited, and distributional considerations are essential. Complementary LCA analyses show environmental competitiveness against cars is unlikely without low-carbon electricity and battery advances \cite{Andre2019LCA}. Embedding UAM within an urban spatial-equilibrium model, \cite{Straubinger2022GoingElectric} links transport with land, labor, and goods markets and finds that welfare and emission outcomes depend on the baseline (gasoline vs. electric cars), with UAM alongside electric cars potentially worsening welfare and raising emissions. Together, these works address transportation-mode comparisons and broader sustainability, land-use, and equity concerns, motivating the presented design approach that integrates vehicle design with operations, economics, and environmental impact.

Beyond eVTOL-specific MDO, earlier transport research has introduced cross-modal performance metrics to compare the efficiency of different modes of transportation. The Gabrielli–von Kármán diagram established fundamental limits of transport performance across speed regimes \cite{Gabrielli1950}, and subsequent updates have extended this framework with modern data and improved figures of merit \cite{Yong2005,Putri2023}. Other studies have focused on eco-efficiency indicators for urban transport \cite{Moriarty2015,Silva2014}, balancing energy use, emissions, and service value. Case studies have analyzed eVTOLs against competing modes in powertrain-level comparisons \cite{Donateo2022}, and in location-specific contexts such as Paris \cite{Hagag2023} and Chicago \cite{Perez2025}. These works provide valuable insights into cross-modal trade-offs and high-level transport efficiency, urban eco-efficiency metrics, and location-specific case studies. A generalizable and integrated metric for systematically benchmarking eVTOLs against the broader transport system within the design process is still missing, motivating the cross-transportation Figure of Merit introduced in this work.

\section{Methodology}
\label{ch:met}

\subsection{Multidisciplinary Design Optimization}
\label{ch:mdo}

The presented eVTOL aircraft design framework uses a Sequential Least Squares Programming (SLSQP) \cite{Kraft1988SLSQP, Kraft1994TOMP} algorithm to solve a set of single-objective, multidisciplinary design optimizations. A multi-start strategy is applied, where the algorithm is run from multiple random initial points, to mitigate sensitivity to local minima. The design variables are normalized to the \([0,\,1]\) range to improve numerical stability \cite{martins2021engineering}. Sensitivities for the gradient-based search are computed using forward finite-difference approximations \cite{gill1983}. Table \ref{tab:opt1} summarizes the optimization problem statement for the four independent and separately run single-objective optimizations of the Figure of Merit, total operation cost, operating profit, and operational global warming potential. A detailed description and remarks on the selection of the optimization approach can be found in Section 8 of the supplementary materials. Figure \ref{fig:xdsm} illustrates the extended design structure matrix \cite{Martins2012} outlining the model integration. The optimization feeds into a system model including mass estimation, environmental impact assessment, cost modeling, and the scoring-based Figure of Merit. Model details are provided in the following sections and supplementary material.

\begin{figure}[ht]
    \centering
    \includegraphics[width=\textwidth]{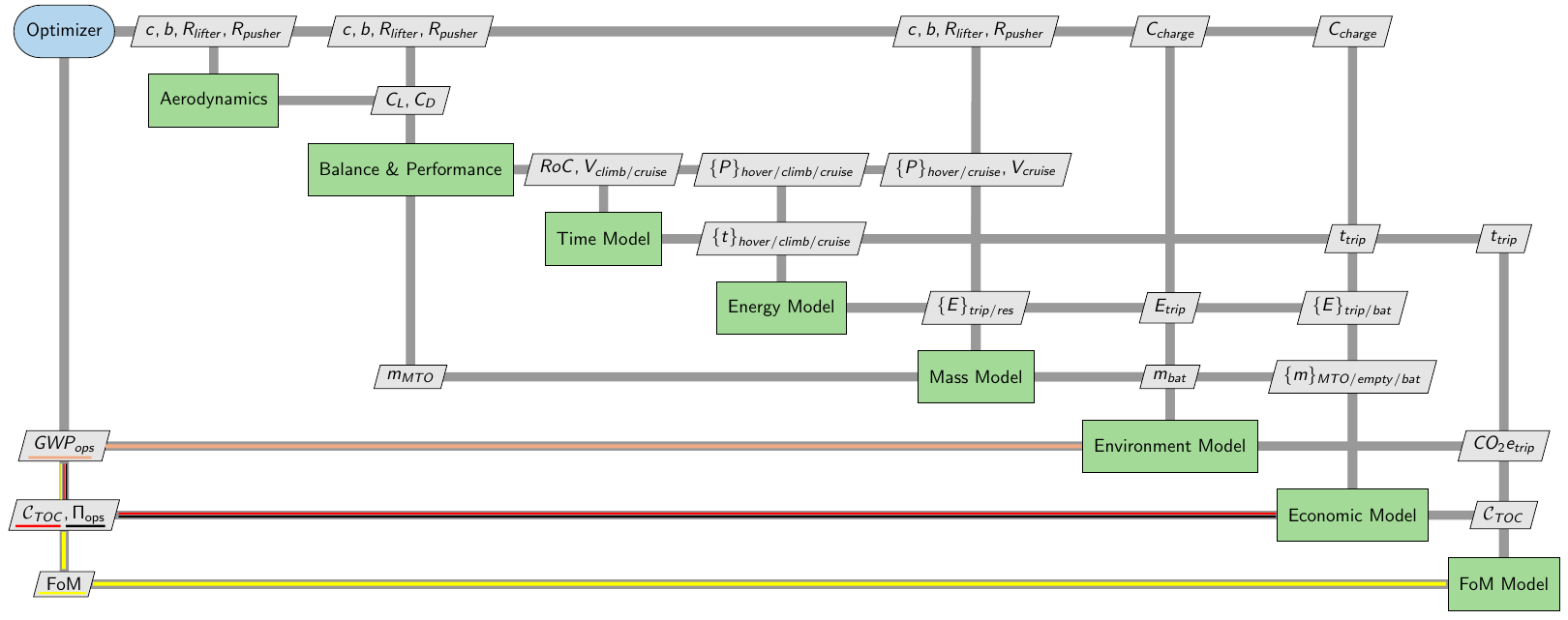} 
    \caption{Extended Design Structure Matrix for single-objective optimization: $FoM_{evtol}$, yellow line; Total Operating Cost $\mathcal{C}_{\text{TOC}}$, red line; Operating Profit $\Pi_{\text{ops}}$, black line; Operational Global Warming Potential $GWP_{ops}$, light brown.}
    \label{fig:xdsm}
\end{figure}

\vspace{-2mm}
\begin{table}[ht]
\centering
\renewcommand{\arraystretch}{1.2}  
\vspace{4pt}
\footnotesize 
\begin{tabular}{llll}
\hline
 & & \textbf{Function/Variable} & \textbf{Description} \\
\hline
\textbf{Objective Type} & objective & max. $FoM_{\text{evtol}}$ & Transportation Figure of Merit (1–10) \\
\textbf{(in separate SO runs)} & & min. $\mathcal{C}_{\text{TOC}}$ & Total operation cost per mission (€) \\
& & max. $\Pi_{\text{ops}}$ & Operating profit (€) \\
& & min. $GWP_{ops}$ & Operational global warming potential (kg CO$_2$e) \\
\hline
\textbf{by varying} & design vars & $R_{\text{lifter}}$ & Lifting rotor radius (m)  \\
& & $R_{\text{pusher}}$ & Pusher rotor radius (m)\\
& & $c$ & Aircraft wing chord (m)\\
& & $b$ & Aircraft wing span (m) \\
& & $C_{charge}$ & Battery charging rate (C-rate) (1/h) \\ 
\hline
\textbf{subject to} & geometry & $b \geq 2(3R_{\text{hover}} + 2d + r_{fus})$ & Wing span must fit hover rotor and fuselage layout (m) \\
& & $15 \geq 2(4R_{\text{hover}} + 2d + r_{fus})$ & Total layout must fit average vertiport width (m) \cite{fuelbirds2023} \\
& regulation & $M_{\text{MTOM}} \leq 5{,}700$ & Max. take-off mass limit (kg) \cite{vtol2024} \\
& noise & $SPL_{\text{hover}} \leq 77$ & Max. noise level at 250 ft (dB(A)) \cite{uberelevate} \\
& rotor speed & $RPM_{\xi} \leq 3{,}000$ & Max. propeller rotational speed per phase (rpm) \\
& airspeed & $V_{\xi} \leq 129$ & Speed limit for climb and cruise phases (m/s) \cite{vtol2024} \\
\hline
\end{tabular}
\vspace{6pt}
\caption{Formulation of the single-objective optimization (SO) problem, repeated for four distinct objectives. $d$ is the required rotor-to-rotor clearance, $r_{fus}$ is the fuselage radius, and ${\xi}$ indexes the flight phases: hover, climb, and cruise.}
\label{tab:opt1}
\end{table}

The design variables of Table \ref{tab:opt1} are bounded to reflect practical design limits and current and near-term feasibility of eVTOL aircraft systems. The wingspan is bounded as \( b \in [6, 15]\,\mathrm{m} \), the wing chord as \( c \in [1, 2.5]\,\mathrm{m} \), the cruise rotor radius as \( R_{\text{cruise}} \in [0.5, 2.5]\,\mathrm{m} \), and the hover rotor radius as \( R_{\text{hover}} \in [0.5, 2.0]\,\mathrm{m} \). The charging rate is bound to \( C_{\text{charge}} \in [1C, 4C] \). A brief classification of the optimization variable bounds is provided in Section \ref{ch:bounds}. A detailed derivation of the wing and vertiport geometry constraints can be found in Section 7 of the supplementary material.

\subsubsection{Mass Estimation}
\label{ch:models}

The maximum take-off mass ($M_{\text{MTOM}}$) consists of the payload mass \( M_{payload} \), the eVTOL empty mass \( M_{empty} \), and the battery mass \( M_{battery} \): 

\begin{equation}
    M_{\text{MTOM}} = M_{payload} + M_{empty} + M_{battery} \leq Limit_{certification}.
    \label{eq:mtom}
\end{equation}

According to the European Union Aviation Safety Agency's (EASA) latest certification specifications for eVTOL, the maximum certifiable mass is limited to \( Limit_{certification} = 5{,}700 \, \text{kg} \) \cite{easa_special_condition_vtol}. \( M_{battery} \) is considered separately from \( M_{empty} \) due to the significant impact the batteries have on the MTOM, and as batteries can be exchanged flexibly if necessary. A detailed description of the mass models is provided in Section 2 of the supplementary material. $M_{battery}$ is calculated using the following equation taking into account the design mission energy requirements and battery energy density \( \rho_{bat} \) Wh/kg, with a factor of 0.64 to account for unusable battery energy, introduced in Section~\ref{ch:powerenergy}. We assume a fixed pack-level energy density of 400 Wh/kg, consistent with mid-term industry targets for advanced lithium-ion chemistries \cite{hollistic, yang2021challenges}. Fixing this parameter allows the analysis to focus on operational and geometric design variables rather than battery technology uncertainty, while reflecting a forward-looking but technically plausible scenario for future UAM deployment \cite{brown2020}.

\begin{equation}
     M_{battery} = \frac{E_{useable}}{e_{usable}} = \frac{E_{trip}+E_{res}}{0.64 \cdot \rho_{bat}}. 
    \label{eq:battery}
\end{equation}

The empty mass $M_{empty}$ is defined as:

\begin{equation}
    M_{empty} = M_{wing} + M_{gear} + M_{rotor} + M_{motor} + M_{fuselage} + M_{systems} + M_{furnish} + M_{crew},
    \label{eq:emptyweight}
\end{equation}

where \( M_{wing} \), \( M_{fuselage} \), \( M_{gear} \), \( M_{systems} \), and \( M_{furnish} \) are estimated according to Raymer \cite{raymer1992aircraft} and Nicolai \cite{weightestimation}. These regression models serve as statistical first-order estimates based on vehicle geometry derived from conventional general aviation aircraft, and may not capture eVTOL-specific structural requirements such as vertical crash loads or extensive composite use. They are applied as conceptual approximations, with higher-fidelity FEM analyses for diverse load cases identified as a potential extension in future work. \( M_{furnish} \) is the predicted weight of furnishings, non-structural components that are necessary to equip the aircraft for the operation and comfort of the crew and passengers.
Crew mass $M_{crew}$ is based on average masses for single pilot operation \cite{easa_passenger_weights_2022}. The rotor mass \( M_{rotor} \) is computed by a radius-based regression adopted from \cite{Martins2023}, intended for early conceptual design. The motor mass \( M_{motor} \) is modeled by a power-based empirical regression \cite{govindarajan2020conceptual}, including a power sizing margin factor of $s=1.5$. This factor scales the required mission power to the available installed power per propulsor to ensure that $P_{\mathrm{available}} \geq P_{\mathrm{required}}$ across all design-mission segments. The modeled $M_{rotor}$ and $M_{motor}$ provide statistical first-order estimates and do not capture detailed design requirements. Higher-fidelity methods may be applied in future work to refine these estimates. \( M_{payload} \) accounts for average mass of passenger \( M_{pax} \) and luggage per passenger \( M_{lug} \)  \cite{easa_passenger_weights_2022}, calculated as:

\begin{equation}
     M_{payload} = (M_{pax}+M_{lug}) \cdot n_{seats} \cdot LF. 
\end{equation}

In this work, we consider a fully loaded 4-seater eVTOL (excl. pilot) as sizing requirement, resulting in a fixed payload of 392.8 kg.

\subsubsection{Aerodynamic Model}

We use a simplified aerodynamic lift and drag coefficient model of the eVTOL wing \cite{Martins2023, Chauhan2019}, and use airfoil data for a NACA2412 airfoil at $Re = 4 \times 10^6$ \cite{drela_xfoil}.  A detailed description of the aerodynamic models is provided in Section 3 of the supplementary material. While this simplified model is convenient for fast optimization and allows for the consideration of multiple subsystems, the same framework can be extended to more complex models in future work. The model for the lift coefficient $C_L$ assumes a linear pre-stall lift curve, using the finite-wing lift coefficient:

\begin{equation}
C_L = \frac{\alpha_{airfoil}}{1 + \frac{\alpha_{airfoil}}{\pi \cdot AR \cdot e}} \cdot \alpha + C_{L_0},
\end{equation}

where \( \alpha_{airfoil} \) is the airfoil lift-curve slope, \( AR \) is the wing aspect ratio, and \( e \) is the Oswald efficiency factor. Total lift is then calculated as:

\begin{equation}
    L = \frac{1}{2} \cdot \rho \cdot V^2 \cdot S \cdot C_L = \frac{1}{2} \cdot \rho \cdot V^2 \cdot S \cdot \left( \frac{\alpha_{airfoil}}{1+\frac{\alpha_{airfoil}}{\pi \cdot AR \cdot e}} \cdot \alpha + C_{L_0} \right).
\end{equation}

The model for the drag coefficient $C_D$ is based on lifting line theory \cite{Chauhan2019}, and assumes that the finite-wing drag coefficient is composed of induced drag \( C_{D_i} \) and parasite drag \( C_{D_{min}} = 0.0397 \) \cite{Martins2023}:

\begin{equation}
C_D = C_{D_{min}} + C_{D_i} = C_{D_{min}} + \frac{C_L^2}{\pi \cdot AR \cdot e}.
\end{equation}

The total drag is then calculated as:

\begin{equation}
D = \frac{1}{2} \cdot \rho \cdot V^2 \cdot S \cdot C_D = \frac{1}{2} \cdot \rho \cdot V^2 \cdot S \cdot \left( 0.0397 + \frac{C_L^2}{\pi \cdot AR \cdot e} \right).
\end{equation}

\subsubsection{Power \& Energy Model}
\label{ch:powerenergy}
The models presented in this subsection are fully derived in Chapter 4 of the supplementary material. The power requirements for hover $P_{hover}$, climb $P_{climb}$, and cruise $P_{cruise}$ are computed using momentum theory derived from force balance equilibrium \cite{rotaru2018helicopter, marchman2021altitude}, considering thrust requirements of the respective flight phase $T_{req}$, rotor diameter $R$, max. take-off mass, air density $\rho$, induced velocity $v_i$, climb angle $\theta$ (assuming zero wind, thus $\theta=\alpha$) and propulsion efficiency $\eta$ \cite{brown2018vehicle, yang2021challenges}. The system efficiencies are described in Section 4.1 of the supplementary material. Due to the conceptual nature of our work, we assume no wing incidence and zero wind effects. Therefore, we consider angle of attack ($\alpha$), pitch angle ($\theta$), and flight path angle ($\gamma$) to be equivalent ($\alpha=\theta=\gamma$). The time requirement for each flight phase is calculated considering the speeds for climb $V_{climb}$ and cruise $V_{cruise}$, derived from force balance equilibrium of the related flight phase \cite{marchman2021altitude}. Thrust angle $\alpha_t$ and pitch angle $\theta$ are assumed to be equivalent to the angle of attack during cruise, $\alpha_{cruise}$, and climb $\alpha_{climb}$, respectively. The hover time is assumed to be fixed at a total of 60 seconds \cite{uberelevate}. This results in the distance-bridging energy requirement for the sizing mission:

\begin{equation}
    E_{trip} = \sum_i P_i \cdot t_i = P_{hover} \cdot t_{hover} + P_{climb} \cdot t_{climb} + P_{cruise} \cdot t_{cruise}.
    \label{eq:energy}
\end{equation}

The maximum usable trip energy in the battery results from the consideration of 30min reserve in cruise power conditions, and unusable energy in the battery. The unusable energy is derived from 10\% ceiling and 10\% floor state-of-charge (SOC), and the assumption that the battery is operated in its end-of-life (EOL) state, defined by a 20\% degradation~\cite{yang2021challenges}, resulting in specific battery energy at EOL: 
\begin{equation}
    e_{BAT_{EOL}} = 0.8 \cdot \rho_{bat}.
\end{equation}

For a given battery energy density $\rho_{bat}$ in Wh/kg, the usable specific energy available for the sizing mission is: 
\begin{equation}
    e_{\text{usable}} = 0.8 \cdot e_{BAT_{EOL}} = 0.64 \cdot \rho_{bat} = e_{trip_{sizing}} + e_{reserve}.
\end{equation}

\subsubsection{Rotor Acoustic Model}

The noise emitted by eVTOL is seen as one of the main challenges for social acceptance and thus the general success of urban air mobility \cite{EASA2021social}. For simplicity and clarity, tonal rotor noise is modeled for hover flight, while broadband noise is modeled for climb and cruise phases. Combined contributions are not considered. The Sound Pressure Level (SPL) in hover, climb, and cruise are considered as constraints in the optimization, as outlined in Table \ref{tab:opt1}. We assume an allowable SPL of 77 dB(A) at an observer distance of 250ft, which is 10dB(A) lower and perceived half as loud as a typical small four-seat helicopter \cite{uberelevate}. This threshold was chosen to align with the noise levels of existing urban road traffic, such as those from medium-sized trucks, which are generally considered acceptable within the ambient noise of residential areas \cite{uberelevate}. To compute the tonal SPL for hovering rotors, we apply the Gutin-Deming model \cite{rotornoise, deming, gutin}. The equations for these models are detailed in Section 7.1 of the supplementary material, including all input variables and constants.

The broadband sound pressure level is based on the model by Schlegel, King, and Mull (SKM) \cite{SKM}, accounting for several key factors influencing noise during the climb and cruise phases. The SPL is primarily determined by a constant, the propeller's effective area, and the blade tip velocity. The model also includes corrections for the propeller's lift coefficient and the observer's distance, which change between the climb (250 ft) and cruise (4000 ft) phases.

While the SKM model provides a practical estimation of broadband rotor SPL based on semi-empirical scaling, its accuracy is limited by simplified aerodynamic assumptions and empirical calibration, and should be complemented by higher-fidelity methods in later design stages. In addition, it should be noted that SPL alone is not entirely representative of public noise exposure, as individual annoyance is also dependent on the physiological perception of loudness. This is quantified by the Day Night Level (DNL) \cite{uberelevate}, and should be monitored and assessed individually at the vertiport sites during operation. The cumulative noise impact of eVTOL fleet operations should be considered in future work, as large scale deployment may require lower SPL targets such as 67dB at 250ft \cite{uberelevate}. This more stringent constraint is not considered in the presented study to avoid over-constraining the model. Further modeling details and assumptions are given in Section 7.1 of the supplementary material.

\subsubsection{Economic Model}

The presented economic model considers a hierarchical cost model structure based on Air Transport Association (ATA) \cite{Faulkner1973} and a revenue model inspired by UberElevate \cite{uberelevate, uberelevate2023}. The operations are based on Uber's assumptions of 280 annual working days per eVTOL with an 8-hour operating window.  While not modeled in this work, real-world operation windows need to be evaluated more closely in future studies in light of regional weather conditions and weather statistics, as it can be assumed that eVTOLs will initially operate exclusively under visual flight rules (VFR), which excludes operations in hazy, low visbility weather conditions \cite{hollistic}. 

A detailed derivation of the cost models and underlying assumptions is provided in Section 5.2 of the supplementary material. Total operating costs per flight (TOC) are defined as the sum of cash operating costs (COC), cost of ownership (COO) and indirect operating costs (IOC). The sum of COC and COO is defined as direct operating costs (DOC):
\begin{equation}
    TOC = COC + COO + IOC, 
\end{equation}

with IOC being a fixed percentage of 22\% of DOC \cite{icao_airlines_operating_costs}. COC are directly related to flight operations taking into account energy cost $C_E$, salary-based crew cost $C_C$, mass-based navigation/ATC cost $C_N$ and maintenance cost $C_M$. $C_M$ is the sum of wrap-rated maintenance cost $C_{wrm}$ according to \cite{brown2018vehicle} and battery replacement cost $C_{MB}$:
\begin{equation}
    COC = C_E + C_C + C_N + C_{wrm} + C_{MB}.
\end{equation}

COO accounts for insurance $C_{ins}$, as 6\% of COC \cite{Mihara2021}, and capital expenditure, as annual depreciation cost $C_{dep}$ \cite{annuity_method_depreciation}:
\begin{equation}
    COO = C_{ins} + C_{dep}.
\end{equation}

The profit model is based on the revenue calculation using a fixed kilometer-based fee \cite{uberelevate}. This fee is based on the mileage rate for trips during the day of London cabs \cite{bettertaxi_london} as a first estimation. Future extensions of this model should consider demand-based revenue modeling to capture more realistic revenue management strategies. The accumulated annual profit $\Pi_{\text{ops}}$ over the number of flight cycles $FC_a$ is then expressed as:  
\begin{equation}
    \Pi_{\text{ops}} = FC_a \cdot (Rev - TOC) = FC_a \cdot (fare \cdot d_{\text{trip}} -TOC).
\end{equation}

While the cost model components are adapted from established frameworks, their novelty in this study lies in their integration with environmental (GWP) and transportation preference (FoM) models within a unified multidisciplinary optimization framework. This enables a consistent assessment of economic, environmental, and operational trade-offs in conceptual eVTOL design. 

\subsubsection{Global Warming Potential Model}

The global warming potential (GWP) considered in this work is derived from the GWP impact of electricity generation and the life-cycle emissions of the batteries used. The GWP impact of the energy consumed is based on the International Energy Agency (IEA) data for Germany \cite{iea_germany, ipcc2014annexiii, worldnuclear_carbon}, assumed as 0.38 kg carbon dioxide equivalent ($CO_2$e) per kWh. This metric serves as a common unit of measure to compare the global warming impacts of different greenhouse gases relative to CO$_2$, allowing for a combined assessment of the total emissions footprint \cite{EPA_GWP}. However, our framework also allows for the application of the energy production GWP of a large number of other regions. The operational energy-based GWP per flight cycle is calculated as:
\begin{equation}
\mathrm{GWP}_{\text{energy,cycle}} = E_{\text{trip}} \cdot \mathrm{GWP}_{\text{energy}},
\label{eq:gwp_cycle}
\end{equation}
where $E_{\text{trip}}$ is the mission energy consumption [kWh] and $\mathrm{GWP}_{\text{energy}}$ is the specific GWP of electricity production [kgCO$_2$e/kWh]. The output $\mathrm{GWP}_{\text{energy,cycle}}$ is given in [kgCO$_2$e].

A simple empirical battery degradation model is used, which is described in Section 6 of the supplementary material. We model the life-cycle of a lithium-ion battery based on the depth of discharge, battery charging rate, and flight phase average battery discharge rate. If we relate the life-cycle of the battery to the operational charging cycles, we can estimate the number of annual required batteries. 
\begin{equation}
N_{\text{bat,year}} = 
\frac{N_{\text{wd}} \cdot T_D}{N_{\text{cycles}} \cdot t_{\text{trip}} \cdot DH},
\label{eq:annual_batt}
\end{equation}
where $N_{\text{cycles}}$ is the available discharge cycle life of the battery, $N_{\text{wd}}$ the number of working days per year, $T_D$ the daily operating time, $t_{\text{trip}}$ the average trip time, and $DH$ the time efficiency ratio defined as $\tfrac{t_{\text{turnaround}}}{t_{\text{trip}}}+1$ (see Section~\ref{ch:bounds}). This first-order estimate does not account for battery swapping strategies or partial-cycle utilization.

We assume a life-cycle GWP of 124.5 kg $CO_2$e per kWh battery capacity \cite{Arshad2022}, based on Nickel-Cobalt-Manganese (NCM) lithium-ion batteries, which are widely used in electric vehicles and increasingly in aviation due to their high energy density and favorable weight-to-performance ratio \cite{hollistic}. The GWP of the battery accounts for raw material extraction and processing, battery manufacturing, transportation, and end-of-life treatment \cite{Andre2019}. The battery life-cycle GWP per flight cycle is calculated as:
\begin{equation}
\mathrm{GWP}_{\text{battery,cycle}} \;=\; 
\frac{N_{\text{batt,year}}}{FC_a} \cdot 
\mathrm{GWP}_{\text{battery}} \cdot E_{\text{battery,design}},
\end{equation}
where $\mathrm{GWP}_{\text{battery}}$ is the specific GWP of battery production [kgCO$_2$e/kWh], 
$E_{\text{battery,design}}$ the design battery capacity [kWh], 
$N_{\text{batt,year}}$ the number of batteries required annually, 
and $FC_a$ the annual number of flight cycles. Further details on the operational metrics are provided in Section 5 of the supplementary material. The total annual operational GWP is calculated as:
\begin{equation}
\mathrm{GWP}_{\text{ops}} = FC_a \cdot 
\left(\mathrm{GWP}_{\text{energy,cycle}} + \mathrm{GWP}_{\text{battery,cycle}}\right)
\end{equation}

It should be noted that the presented GWP assessment does not include the full life-cycle of the eVTOL aircraft itself, such as structural components, manufacturing, maintenance, or end-of-life disposal. Nonetheless, the presented framework and analysis can be expanded to include this aspect in future work, e.g. based on the work in \cite{Andre2019}.

\subsubsection{Transportation Figure of Merit}
\label{ch:fom}

We introduce the Transportation Figure of Merit (FoM) to analyze and compare the attractiveness of eVTOL versus other transportation alternatives, and to examine when and why a mode of transportation can be considered competitive in relation to specified objectives. This metric is a stakeholder-centric index that aims to quantify the utility of each transport option (see Table \ref{tab:fom_base}) by time, cost, and $CO_2e$ emissions in a single rating. This allows for comparing the designed eVTOL aircraft with alternative means of transportation over equivalent travel distance. In the context of our analysis, we refer to the sizing mission of 70 km (see Figure \ref{fig:sizing_mission}). The sizing mission distance is the straight-line distance traveled by the eVTOL. To account for real traffic patterns, we use the mode-specific circuity factor to compute the true distance traveled \cite{uberelevate}, the ratio between the actual distance traveled and the straight-line path. This takes into account detours caused by road networks, flight routes, or rail topology to provide a more realistic travel estimate. In this context, the load factor indicates the percentage of available seating capacity that is filled by paying customers \cite{DEVRIENDT2009337}.

For each transport mode $j$, the values for cost, $CO_2e$, and travel time ($x_{j,k}$ for $k \in \{1 = \text{cost},\ 2 = \text{CO}_2,\ 3 = \text{time}\}$) are obtained by applying the base metrics presented in Table \ref{tab:fom_base} to the defined trip distance. The three criteria ($k \in {1, 2, 3}$) were selected due to their availability, quantifiability, and relevance in both stakeholder-centric and policy-related transport assessment. Metrics such as reliability, accessibility, availability, safety, and travel demand (e.g., mode-specific trip density or service coverage) also provide valuable insights, but are not included in our analysis and should be integrated into future extensions of this framework.

\begin{table}[ht]
    \centering
    \begin{tabular}{lccccc}
        \textbf{Vehicle (Load factor), $j$} & \textbf{Ø-V (km/h)}  & \textbf{Ø Cost/skm}  & \textbf{Ø $CO_2e$/skm}  & \textbf{Circuity (-)}  & \textbf{Source} \\ 
        \hline \hline
        Gasoline Car (100\%) & 85 & 0.023 & 0.031 & 1.3 & \cite{ADAC_Golf, Statista_Super10_Germany, Comcar_CO2_Litre} \\
        Gasoline Car (26\%) & 85 & 0.090 & 0.120 & 1.3 &  \\
        Gasoline Car (20\%) & 85 & 0.117 & 0.157 & 1.3 &  \\ \hline
        Diesel Car (100\%) & 85 & 0.017 & 0.026 & 1.3 & \cite{ADAC_VWGolf_diesel, Cambridge_EnergyUse, Statista_Diesel_Germany, Michelin_CO2_Emissions} \\
        Diesel Car (26\%) & 85 & 0.064 & 0.099 & 1.3 & \\
        Diesel Car (20\%) & 85 & 0.083 & 0.128 & 1.3 & \\ \hline
        Electric Car (100\%) & 85 & 0.021 & 0.013 & 1.3 & \cite{EVDB_Tesla_ModelY, Statista_EVCharging_Germany, ScienceDirect_2015_CO2} \\
        Electric Car (26\%) & 85 & 0.081 & 0.050 & 1.3 & \\
        Electric Car (20\%) & 85 & 0.105 & 0.065 & 1.3 & \\ \hline
        Public Bus (diesel) (100\%) & 64 & 0.06 & 0.013 & 1.6 & \cite{Flixbus_Atmosfair_WTW, Cambridge_EnergyUse, Michelin_CO2_Emissions, Statista_Diesel_Germany} \\
        Public Bus (diesel) (60\%) & 64 & 0.104 & 0.022 & 1.6 & \\ \hline
        Public Train (electric) (100\%) & 99 & 0.2 & 0.007 & 1.2 & \cite{ChemEurope_FuelEfficiency, Siemens_ICE4} \\
        Public Train (electric) (50\%) & 99 & 0.402 & 0.012 & 1.2 & \\ \hline
        Airplane, jet (100\%) & 74$^{(*)}$ & 0.46 & 0.198 & 1.05 & \cite{Skytanking_JetFuel, IATA_CarbonOffset_FAQ} \\
        Airplane, jet (79.6\%) & 74$^{(*)}$ & 0.579 & 0.249 & 1.05 & \\ \hline
        Bicycle (100\%)  & 18.8 & -0.491 $^{(**)}$ & 0.0 & 1.28 & \cite{Sustrans_CyclingValue, ScienceDirect_AustralianTransport_1999} \\ \hline
    \end{tabular}
    \vspace{6pt}
    \caption{Transportation performance metrics for average travel velocity, cost, emission expenditure, and circuity of conventional transportation modes, used as baseline for the Transportation Figure of Merit (FoM) comparison. Sources apply to all load factor scenarios within each vehicle class. $^{(*)}$ Accounts for an additional two hours for security and boarding procedures as recommended by airlines. $^{(**)}$ Negative cost as considers personal health impact (skm = seat-kilometer).}
    \label{tab:fom_base}
\end{table}

Each criterion is normalized into a rating between 1 and 10 using Eq. (\ref{eq:rating}), where the transport mode with the best performing metric receives a score of 10, and the worst performing one is rated 1. All other results are interpolated proportionally in between. Let $x_{j,k}$ denote the value of criterion $k$ for vehicle $j$. The normalized rating $R_{j,k}$ is computed as:

\begin{equation}
    R_{j,k} = \frac{(x_{j,k} - X^{(k)}_{\min}) - 10 \cdot (x_{j,k} - X^{(k)}_{\max})}{X^{(k)}_{\max} - X^{(k)}_{\min}},
    \label{eq:rating}
\end{equation}

where $X^{(k)}_{\min} = \min_j x_{j,k}$ and $X^{(k)}_{\max} = \max_j x_{j,k}$ are the minimum and maximum values of criterion $k$ across all vehicles. The overall FoM is subsequently calculated by weighting each rating to quantify the combined utility, with $\sum_{k=1}^{3} w_k = 1$. For vehicle $j$, with the normalized rating $R_{j,k}$ with respect to criterion $k$, the FoM defined as: 

\begin{equation}
    \text{FoM}_j = \sum_{k=1}^{3} w_k \cdot R_{j,k}
    \label{eq:fom}
\end{equation}

For our analysis and the absence of real-world stakeholder preferences, we assume a stakeholder who weights each criterion $k$ equally, to provide a transparent and balanced consideration within the presented work as $w_k = \frac{1}{3} \ \forall \, k$. Future studies may investigate a variation in the weightings to explore the behavior of stakeholders at particular cost, time, or $CO_2e$ sensitivity in more depth to reflect real-world preferences. It should be noted that the normalized ratings are sensitive to the composition of the transport mode set, adding or removing a mode shifts the reference bounds and may alter individual scores.

\subsection{Optimization Bounds}
\label{ch:bounds}

The design variables defined in Table \ref{tab:opt1} in Section \ref{ch:mdo} are bounded to reflect the current feasibility and medium-term technological potential of eVTOL systems. As described in Table \ref{tab:scenario_definitions}, the lower limits correspond to restrictive assumptions based on existing technology and regulatory constraints, while the upper limits represent projected assumptions consistent with expected advances, especially in battery technology, and extended design capabilities. The interpretations of the boundaries are used in the further analysis to derive regulatory and operational conditions for the optimized design and determine appropriate recommendations.

\begin{table}[ht]
\renewcommand{\arraystretch}{1.5}
\footnotesize
\centering
\begin{tabular}{m{3.4cm}|m{5.7cm}|m{5.7cm}}
 & \textbf{Lower bound} & \textbf{Upper bound} \\
\hline
\textbf{Wing span $b$} 
& 6\,m: Compact wing suitable for highly constrained urban settings. 
& 15\,m: Reflects urban space limitations and integration challenges with existing urban infrastructure \cite{helipaddy_wills}. \\
\hline
\textbf{Wing chord $c$} 
& 1.0\,m: Narrow chord consistent with compact, lightweight designs. 
& 2.5\,m: Wide chord accommodates structural and aerodynamic performance at higher loads. \\
\hline
\textbf{Rotor radius $R$} 
& 0.5\,m: Aligns with industry
benchmarks and UAM constraints, offering a realistic framework for current designs (industry comparison in supplementary material Section \ref{ch:sup}).
& 2.5\,m: Allows exploration of scenarios with fewer spatial and noise constraints,
suitable for suburban or rural eVTOL applications. \\
\hline
\textbf{Charging rate $C_{charge}$} 
& 1C: Assumes slow charging, offering a conservative scenario, especially in urban areas with limited fast-charging. 
& 4C: Supports rapid charging, enabling faster service cycles as intended by various UAM providers \cite{hollistic}. \\
\hline
\end{tabular}
\vspace{6pt}
\caption{Optimization bounds: lower bounds represent conservative, near-term feasibility; upper bounds reflect forward-looking technological and infrastructural assumptions.}
\label{tab:scenario_definitions}
\end{table}

The charging $C$-rate directly influences the modeled turnaround time in eVTOL operations. For simplicity, the leg time $t_{\mathrm{leg}}$ is expressed as the sum of flight time and turnaround time, where the flight time is a function of the vehicle velocity in different flight phases, the trip distance, and the hover duration:  

\begin{equation}
    t_{\mathrm{flight}} = f\left(V_{\xi}, D_{\mathrm{trip}}, t_{\mathrm{hover}}\right), \quad \xi \in \{\mathrm{climb}, \mathrm{cruise}\},
\end{equation}

and the turnaround time is modeled as a function of the charging rate and the depth-of-discharge (DoD), taking into account the battery charging time:

\begin{equation}
    t_{\mathrm{turnaround}} = f\left(C_{\mathrm{charge}},  \mathrm{DoD}\right).
\end{equation}

Boarding and disembarking times are not taken into account due to the relatively low passenger throughput per aircraft compared to conventional civil aviation ground procedures. A more detailed modeling of the turnaround process should be considered in future work. Section 5.1 of the supplementary material provides the exact formulation of the underlying models.

\subsection{Sizing Mission}

\begin{figure}[H]
    \centering
    \begin{subfigure}[b]{0.55\textwidth}
        \centering
        \includegraphics[width=\textwidth]{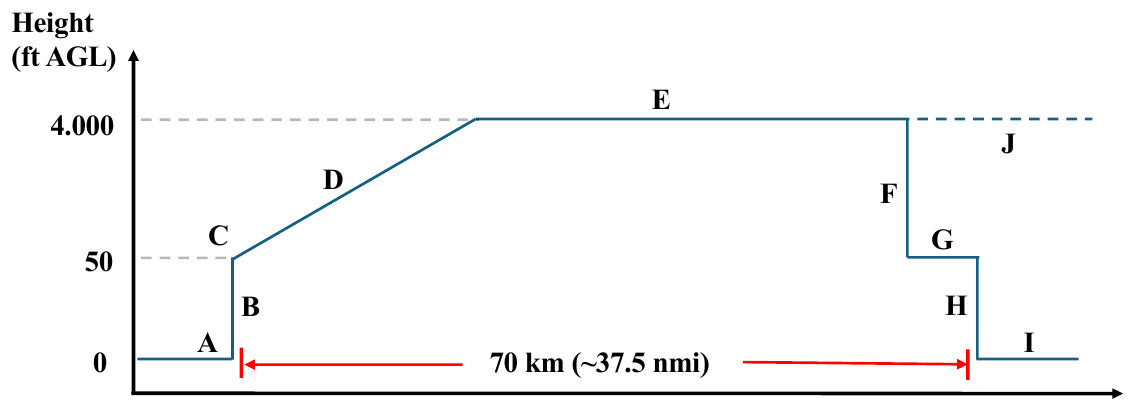}
        \caption{Sizing mission profile.}
        \label{fig:profile}
    \end{subfigure}
    \hfill
    \begin{subfigure}[b]{0.4\textwidth} 
        \centering
        \scriptsize
        \begin{tabular}{cl}
            \hline\hline
            \textbf{Segment} & \textbf{Description} \\
            \hline \hline
            A & Take-off \\
            B & Vertical climb to 50ft (30s hover) \\
            C & Transition \\
            D & Climb \\
            E & Cruise at 4,000ft \\
            F & No credit descent \\
            G & Transition \\
            H & Vertical descent from 50ft (30s hover) \\
            I & Landing \\
            J & 20 min reserve \\
            \hline\hline
        \end{tabular}
        \caption{Sizing mission segments.}
        \label{tab:sizing_mission_segments}
    \end{subfigure}
    \caption{Sizing mission: (a) profile and (b) segments.}
    \label{fig:sizing_mission}
\end{figure}

This study adopts the sizing mission profile proposed by National Aeronautics and Space Administration (NASA) \cite{patterson2018proposed}, which is characterized by a 37.5 nautical mile ($\sim$70 km) travel distance and a cruise altitude of 4.000ft above ground level (AGL), shown in Figure~\ref{fig:sizing_mission}. This widely accepted mission profile aligns with established research (\cite{silva2018}, \cite{patterson2018proposed}), enhancing the comparability of the presented findings within academic and industry contexts. 

The use of these parameters reflects realistic operational conditions for urban and regional air mobility, including typical city-to-city or intra-urban flights. By incorporating key operational phases such as vertical climb, cruise climb and cruise, this mission profile provides comprehensive insights into eVTOL performance, including energy consumption and battery requirements. The modeling of the transition phase is not included in this study to reduce complexity, and should be addressed in future high-fidelity analyses to capture transient aerodynamic effects, control requirements, and potential impacts on energy consumption and noise generation. Additionally, a 20-minute cruise reserve is applied in accordance with the Federal Aviation Administration (FAA) Special Federal Aviation Regulation (SFAR) requirements \cite{FAA2024poweredlift}.

\subsection{Analysis Workflow}
\label{ch:scenarios}

The presented analysis builds on the MDO framework introduced in Section \ref{ch:mdo}. The four objectives are optimized independently in separate single-objective runs, subject to a constant set of defined design variables, bounds, and constraints: (1) maximum annual profit, (2) minimum operating cost per trip, (3) minimum annual global warming potential (GWP), and (4) maximum transportation figure of merit (FoM). This setup ensures comparability across objectives while allowing each optimization to highlight a distinct design priority. The outcome of each single-objective optimization is an optimal conceptual eVTOL configuration and operations setup, tailored to its specific design priority. In Section \ref{ch:results}, these designs are first discussed individually to identify how cost, profit, emissions, or user utility shape the resulting aircraft geometry, performance, and operations. 
Subsequently, the results are compared across objectives to expose trade-offs between economic, environmental, and operational considerations. 
In Section \ref{ch:discussion}, the obtained designs are benchmarked against external industry requirements \cite{uberelevate, uberelevate2023}. By contrasting the resulting optima, we gain insights into potential pathways toward balanced trade-off designs that reconcile competing stakeholder objectives and support the alignment of technical capabilities with anticipated operational environments.

\section{Results}
\label{ch:results}

\subsection{Profit-Maximized eVTOL Design}

\begin{figure}[p]
    \centering
    \includegraphics[width=0.97\textwidth]{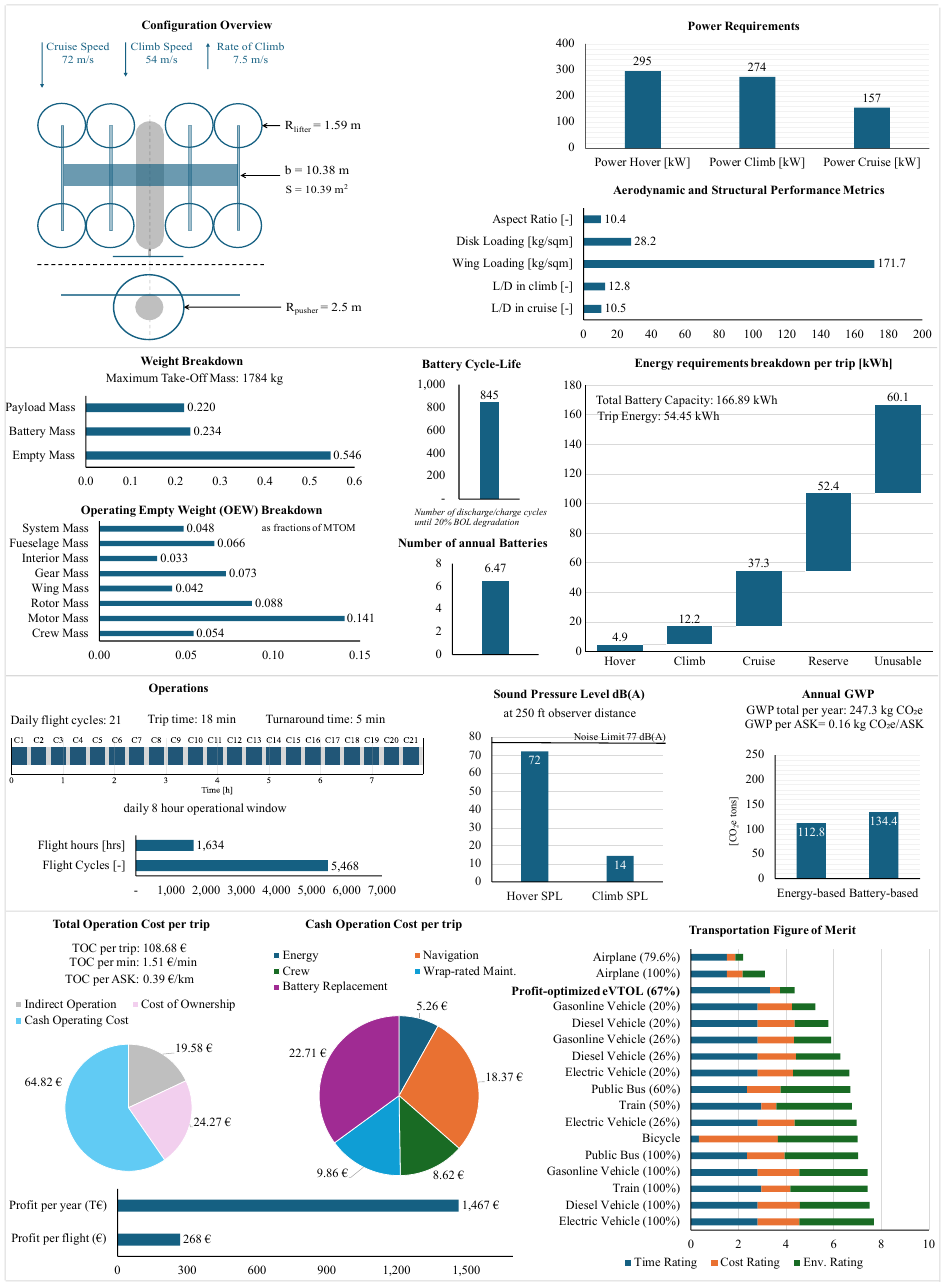} 
    \caption[Optimization results of profit-maximizing conceptual eVTOL design.]{Optimization results of profit-maximizing conceptual eVTOL design: aircraft design, battery behavior, operating window usage, global warming potential, operating costs and cross-transportation comparison.}
    \label{fig:combined_profit_comparison}
\end{figure}

The profit-maximizing eVTOL design, illustrated in Figure \ref{fig:combined_profit_comparison}, was obtained after 8 iterations and 58 function evaluations, with a runtime of 1.36 seconds, and achieves an annual operating profit of €1.47 million. The optimizer converged to a medium wing span of 10.4 m with a slender chord of approximately 1.0 m, a cruise propeller radius of 2.50 m, and moderate hover rotors of 1.59 m, thus remaining within the 15 m vertiport diameter constraint while balancing aerodynamic efficiency and mass. This configuration enables a cruise speed of 72 m/s at a lift-to-drag ratio of 10.5, leading to an 18-minute trip time. The maximum take-off mass is 1,784 kg with an empty-mass fraction of 0.546 and a battery fraction of 0.23, within the typical range for eVTOL designs \cite{hollistic}. As the energy density is assumed as 400 Wh/kg at pack-level, the required design capacity of 166.9 kWh translates into a battery mass of 417 kg.

The moderate-aspect-ratio wing and large cruise propeller radius of 2.50 m support efficient high-speed cruise at 72 m/s, resulting in a trip energy of 54.5 kWh and a short airborne segment of 18 minutes. Combined with the maximum charging rate of 4C, which results in a turnaround time of 5 minutes, this yields a total block time of 23 minutes per cycle, allowing more than 5,400 annual cycles and approximately 1,600 annual flight hours. Profitability is determined by the combined effect of per-flight margin and annual vehicle utilization, both of which are maximized by the selected design. This utilization translates to approximately 21 flights per day, which is only feasible under the assumption of sufficient demand, a factor that should be examined in future demand-supply modeling to extend this work. 

Although battery replacements account for approximately one fifth of the total operating cost of €108 per flight, the high number of daily cycles enabled by short block times more than compensates for this penalty, making annual utilization the dominant driver of profit. Energy cost per trip is €5.26 (5\% of TOC), crew cost is €8.62 (8\% of TOC), and time-based maintenance amounts to €9.86 (9\% of TOC). Navigation charges are €18.37 (17\% of TOC) and ownership costs €24.27 (22\% of TOC), both directly scaling with MTOM. The major penalty of this design arises from the battery management, allowing fast charging times at the cost of accelerated cycle wear, reducing the battery lifetime to about 845 charging cycles and resulting in 6.5 battery replacements per year. With a unit pack cost of €19,192 (115 €/kWh, \cite{Mihara2021}), this corresponds to €22.7 per flight or 21\% of TOC. As shown in Table 12 of Section 7 of the supplementary material, the profit-maximized design is primarily limited by geometric constraints rather than acoustic margins, with the optimizer driving the wingspan to its minimum allowable limit to reduce structural mass. This results in a tightly packaged configuration where the wing length is strictly dictated by the physical clearance required for the hover rotors and fuselage layout. While the design achieves a noise constraint margin of 5 dB(A) (72 dB(A) vs. a 77 dB(A) limit), the very high frequency of operations implied by this design would amplify community noise exposure, which is not yet accounted for in the present study and should be considered in future operational assessments. 

From an environmental impact perspective, the design produces 247.3 tons of CO$_2$e per year, corresponding to 0.16 kg CO$_2$e per available seat-kilometer (151.34 kg CO$_2$e per flight hour). Of this, 46\% is attributed to electricity generation for battery usage and 54\% to battery replacement impact, highlighting that the high battery turnover rate of 6.5 replacements per year is the dominant emissions driver in this configuration. This translates to 58 gasoline-powered passenger vehicles or 218 electric-powered passenger vehicles driven for one year \cite{EPA2025_GHG_Equivalencies}. In cross-transportation comparison, the design scores $R_{EVTOL_{time}}=10$ for time but only $R_{EVTOL_{CO_{2}e}}=1.82$ for emissions and $R_{EVTOL_{cost}}=1.27$ for cost, yielding a composite FoM of 4.36. This places the eVTOL above only conventional aircraft and below gasoline internal combustion vehicles at 20\% load factor, highlighting that although the design is highly profitable from the operator's point of view, it remains disadvantaged when compared to ground transport modes under equal weighting of time, cost, and environmental impact.

\paragraph{Key Takeaways:}
\begin{itemize}
    \item High annual utilization ($\sim$5,469 cycles) is the primary profit driver, enabled by aggressive 4C charging for short turnarounds and high cruise speeds (72 m/s) to minimize block time.
    \item The design accepts a significant maintenance penalty, with battery replacements occurring 6.5 times per year (21\% of TOC), yet the volume of daily flights (21 flights) ensures overall high profitability.
    \item Despite economic success, the design faces significant environmental challenges, scoring poorly in emission-based comparative metrics primarily due to high battery throughput.
\end{itemize}

\vspace{-6pt}

\subsection{Cost-Minimized eVTOL Design}
\vspace{-4pt}

\begin{figure}[p] 
        \centering
        \includegraphics[width=0.97\textwidth]{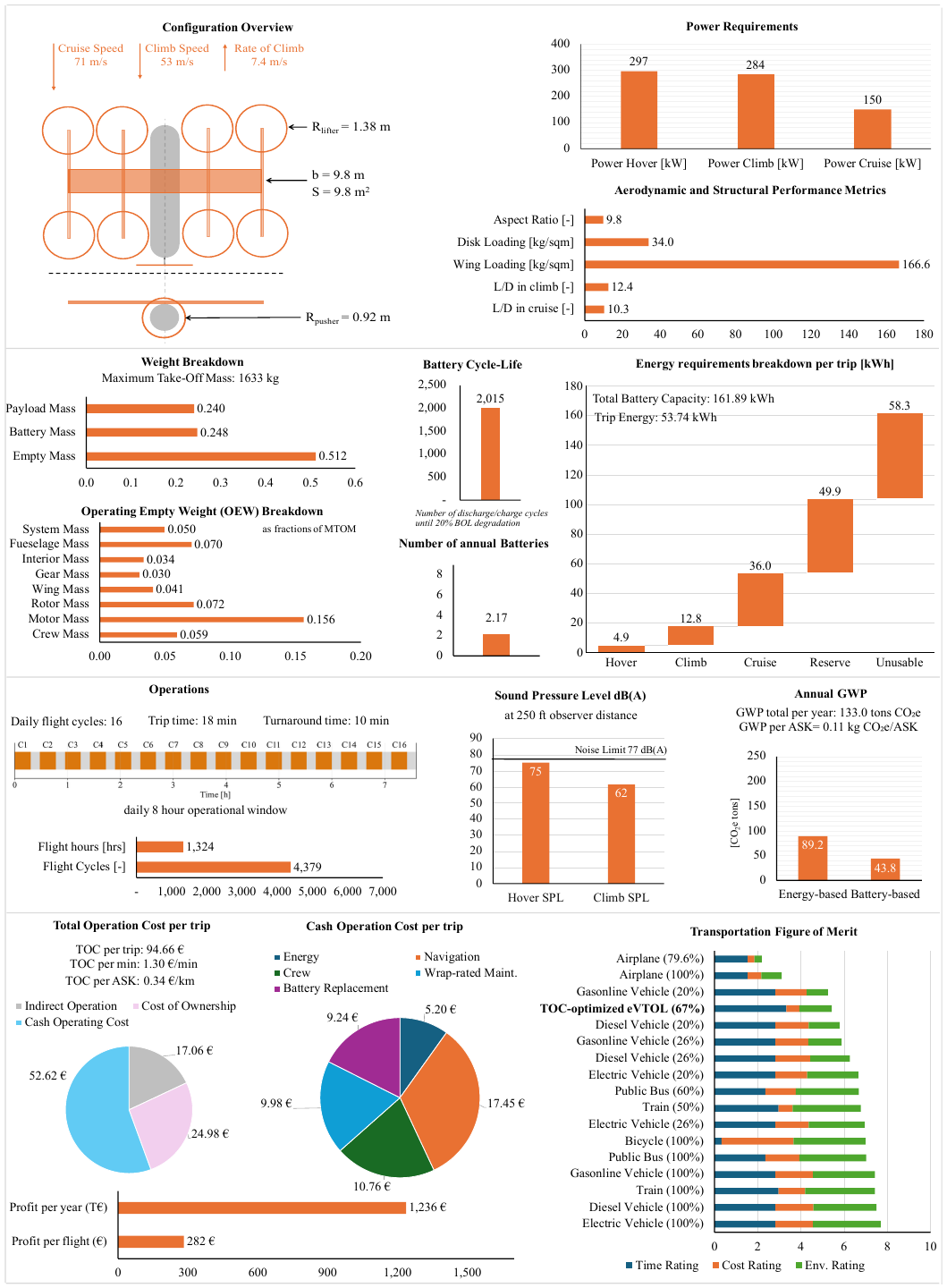}
        \caption[Optimization results of TOC-minimizing conceptual eVTOL design.]{Optimization results of TOC-minimizing conceptual eVTOL design: aircraft design, battery behavior, operating window usage, global warming potential, operating costs and cross-transportation comparison.}
        \label{fig:combined_TOC_comparison}
\end{figure}

The TOC-minimizing eVTOL design, illustrated in Figure \ref{fig:combined_TOC_comparison}, was obtained after 9 iterations, 78 function evaluations, with a runtime of 1.17 seconds. The optimizer converged to a wingspan of 9.80 m with a chord of 1.0 m, a relatively small cruise propeller radius of 0.92 m, and hover rotors of 1.38 m, resulting in a maximum take-off mass of 1,633 kg. The empty-mass fraction is 0.51, with a battery fraction of 0.25, both in typical ranges for eVTOL aircraft \cite{hollistic}. At a pack-level energy density of 400 Wh/kg, the design requires 162 kWh of capacity, corresponding to a battery mass of 405 kg. Cruise speed is 71 m/s with a lift-to-drag ratio of 10.3, leading to a trip time of 18.1 minutes on the 70 km sizing mission. The configuration combines small-sized propeller radii and reduced mass fractions with a relatively low MTOM, reflecting the prioritization of cost reduction over utilization throughput and configuration complexity. The constraints for wing-rotor geometry and climb RPM are active (see Table 12 in the supplementary material), indicating that the optimizer balances wing length and rotor radius to reduce drag and maximize rotor performance within the RPM boundaries. 

Operationally, the charging rate is moderate at 1.9C, resulting in a turnaround time of 10.4 minutes and a block time of 28.5 minutes. This limits daily utilization to 17 flights and 4,379 annual cycles, equivalent to 1,324 annual flight hours. Profitability results from lower utilization offset by significantly reduced per-flight costs. Energy consumption per trip is 53.7 kWh. The depth-of-discharge of 0.33 together with an average discharge C-rate of 1.1 1/h extends the battery life to 2,015 cycles, resulting in 2.2 battery replacements per year and directly impacts recurring maintenance costs. 

The cost structure reflects these operational adjustments. Energy cost per trip is €5.20 (5.5\% of TOC), annual salary-based crew cost is €10.76 (11\% of TOC), and time-based maintenance is €9.98 (11\% of TOC). Navigation charges, at €17.45 (18\%), remain the highest single cost element, together with cost of ownership at €24.98 (26\%), both scaling with MTOM. Battery-related costs are €9.24 per trip, corresponding to about 10\% of total operating costs. Overall, total operating cost is minimized to €94.7 per flight, corresponding to €0.34 per seat-kilometer. Despite lower annual utilization, annual profit reaches €1.24 million, highlighting that cost savings can maintain economic viability even when throughput is reduced. 

The TOC-optimized design produces the highest hover noise level among the four optimized configurations at 75 dB(A), leaving a margin of only 2 dB(A) to the 77 dB(A) constraint. This is a direct consequence of the smaller rotor radii (1.38 m hover, 0.92 m cruise) relative to the other configurations, which increases disk loading and elevates tonal noise despite the moderate MTOM. Annual operational GWP achieves 133 tons CO$_2$e, comparable to 31 gasoline-powered passenger vehicles or 117 electric-powered passenger vehicles driven for one year \cite{EPA2025_GHG_Equivalencies}. On a per-flight basis, emissions amount to 30 kg CO$_2$e (100.45 kg CO$_2$e per flight hour), of which 67\% are related to electricity generation and 33\% to battery replacements. The limited number of battery replacements reduces the environmental burden of energy storage systems, making environmental performance more favorable. 

In cross-modal comparison, the TOC-minimized eVTOL achieves a balanced Figure of Merit. With a trip time of 18.1 minutes, it scores $R_{EVTOL_{time}}=10$ for time. Cost performance remains low at $R_{EVTOL_{cost}}=1.81$, while emissions performance reaches $R_{EVTOL_{CO_{2}e}}=4.51$ due to limited battery turnover. The overall FoM is 5.44, placing this design closer to ground-based electric and diesel vehicles and well above conventional aviation, which indicates improved competitiveness in a broader transport context.

\paragraph{Key Takeaways:}
\begin{itemize}
    \item The design prioritizes component longevity by utilizing a moderate 1.9C battery charging, resulting in a block time of 28.5 minutes and 17 daily flights.
    \item Significant cost reduction is achieved through extended battery life (2,015 cycles), which limits replacements to 2.2 per year and reduces battery-related costs to 10\% of the TOC.
    \item Minimizes total operating costs to €94.7 per flight while maintaining noise levels at 75 dB(A) and achieving a balanced Figure of Merit (5.44) through reduced energy and maintenance throughput.
\end{itemize}

\subsection{GWP-Minimized eVTOL Design}
\begin{figure}[p] 
    \centering
    \includegraphics[width=0.97\textwidth]{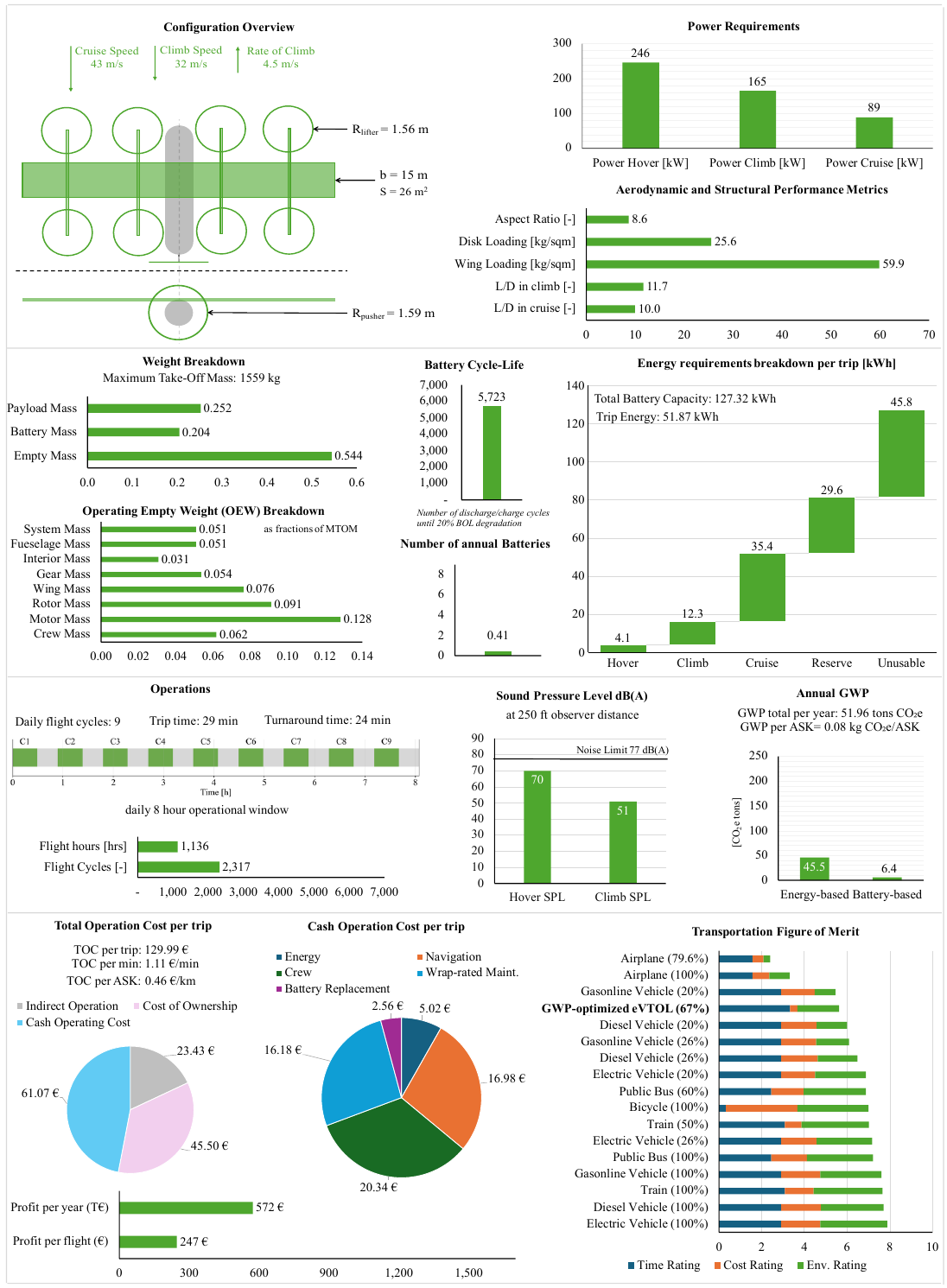}
    \caption[Optimization results of GWP-minimizing conceptual eVTOL design.]{Optimization results of GWP-minimizing conceptual eVTOL design: aircraft design, battery behavior, operating window usage, global warming potential, operating costs and cross-transportation comparison.}
    \label{fig:combined_gwp_comparison}
\end{figure}

The GWP-minimizing eVTOL design, illustrated in Figure \ref{fig:combined_gwp_comparison}, was obtained after 6 iterations and 0.66 seconds of runtime. The optimization yields a wingspan of 15 m with a chord of 1.74 m, a cruise propeller radius of 1.59 m, and hover rotors of 1.56 m. This configuration results in a maximum take-off mass of 1,559 kg, with an empty-mass fraction of 0.54 and a battery fraction of 0.20. At a pack-level energy density of 400 Wh/kg, the design requires 127 kWh of capacity, corresponding to a battery mass of 318 kg. Cruise speed is computed as 43 m/s with a lift-to-drag ratio of 10, leading to a trip time of 29.4 minutes. The combination of large wing span and rotor radii with a relatively low MTOM reflects a prioritization of energy efficiency to minimize indirect operating emissions. The optimizer drives the wingspan to its upper bound of 15 m, constrained by the vertiport footprint, while exhibiting considerable margins across acoustic and weight-based constraints (see Table 12 in the supplementary material), confirming that the active constraint is spatial rather than structural or acoustic.

The optimizer converges to the lower charging bound of 1C, resulting in an extended turnaround time of 24.4 minutes and a block time of 53.9 minutes per trip, constraining daily utilization to 8.9 flights and 2,317 annual cycles, equivalent to 1,136 flight hours. Operating at a cruise speed of 43 m/s results in cruise power of 89 kW, the lowest of all designs. Energy demand per trip is 51.9 kWh, distributed across hover, climb, and cruise phases, with a depth-of-discharge of 0.41 and an average discharge C-rate of 0.83. Annual GWP is reduced primarily through the coupled limitation of energy and battery throughput at the annual operational scale, as the conservative charging strategy and low cruise speed together limit annual flight cycles to 2,317, reducing total annual energy consumption and battery replacement frequency simultaneously. These values indicate relatively gentle battery management, extending battery life substantially to 5,723 charging cycles, resulting in fewer than one battery replacement per year. This very low replacement frequency directly reduces the emission burden associated with battery production and disposal, as well as battery-related replacement costs \cite{Arshad2022}. 

The cost structure is shaped by extended flight and turnaround times. Energy costs per trip are €5.02 (3.9\% of TOC),  navigation charges €16.98 (13\% of TOC), and crew costs €20.34 (16\% of TOC), becoming a dominant cost driver due to extended flight times. Time-based maintenance adds €16.18 per trip (12\% of TOC). Battery-related costs are lowest with €2.56 per trip, representing 2\% of total operating costs. Ownership costs are €45.50 per trip (35\%), keeping total operating cost at €130 per flight, or €0.46 per seat-kilometer respectively. Despite the low frequency of operations, annual profit remains positive at €0.57 million. 

The large lifting surfaces and gentle battery use contribute to both favorable noise margins and low environmental impact. The design remains below the hover noise limit at 70 dB(A). Annual operational GWP is 51.96 tons CO$_2$e, comparable to 12 gasoline-powered passenger vehicles or 46 electric-powered passenger vehicles driven for one year \cite{EPA2025_GHG_Equivalencies}. On a per-flight basis, emissions are 22.4 kg CO$_2$e (45.74 kg CO$_2$e per flight hour), of which 88\% are attributable to electricity generation and 12\% to battery replacements, consistent with the low annual battery replacement rate of 0.41 packs per year. In cross-modal comparison, with a trip time of 29.4 minutes, this design scores $R_{EVTOL_{time}}=10$ for travel time. Cost performance reaches the lowest possible rating at $R_{EVTOL_{cost}}=1$, reflecting the highest total operating cost among the optimized designs at €130 per flight, driven by time-intensive crew and ownership costs. Emissions performance reaches $R_{EVTOL_{CO_{2}e}}=5.95$ due to very low battery turnover and efficient flight. The overall FoM is 5.65, reflecting the design’s strong environmental profile in relation to its operational performance.

\paragraph{Key Takeaways:}
\begin{itemize}
    \item Achieves the lowest environmental impact (46 kg CO$_2$e per flight hour) through the combination of low cruise power at 43 m/s and a conservative 1C charging strategy. 
    \item Employs a conservative 1C charging rate that extends battery life to 5,723 cycles. This reduces battery replacements to less than once per year, making replacements responsible for only 12\% of the per-flight GWP and 2\% of the per-flight TOC.
    \item The operational profile significantly limits utilization to 9 flights per day, causing time-related costs, specifically maintenance (12\% of TOC) and crew (16\% of TOC), to become the dominant cost drivers.
\end{itemize}

\subsection{FoM-Maximized eVTOL Design}

\begin{figure}[p]
    \centering
    \includegraphics[width=0.96\textwidth]{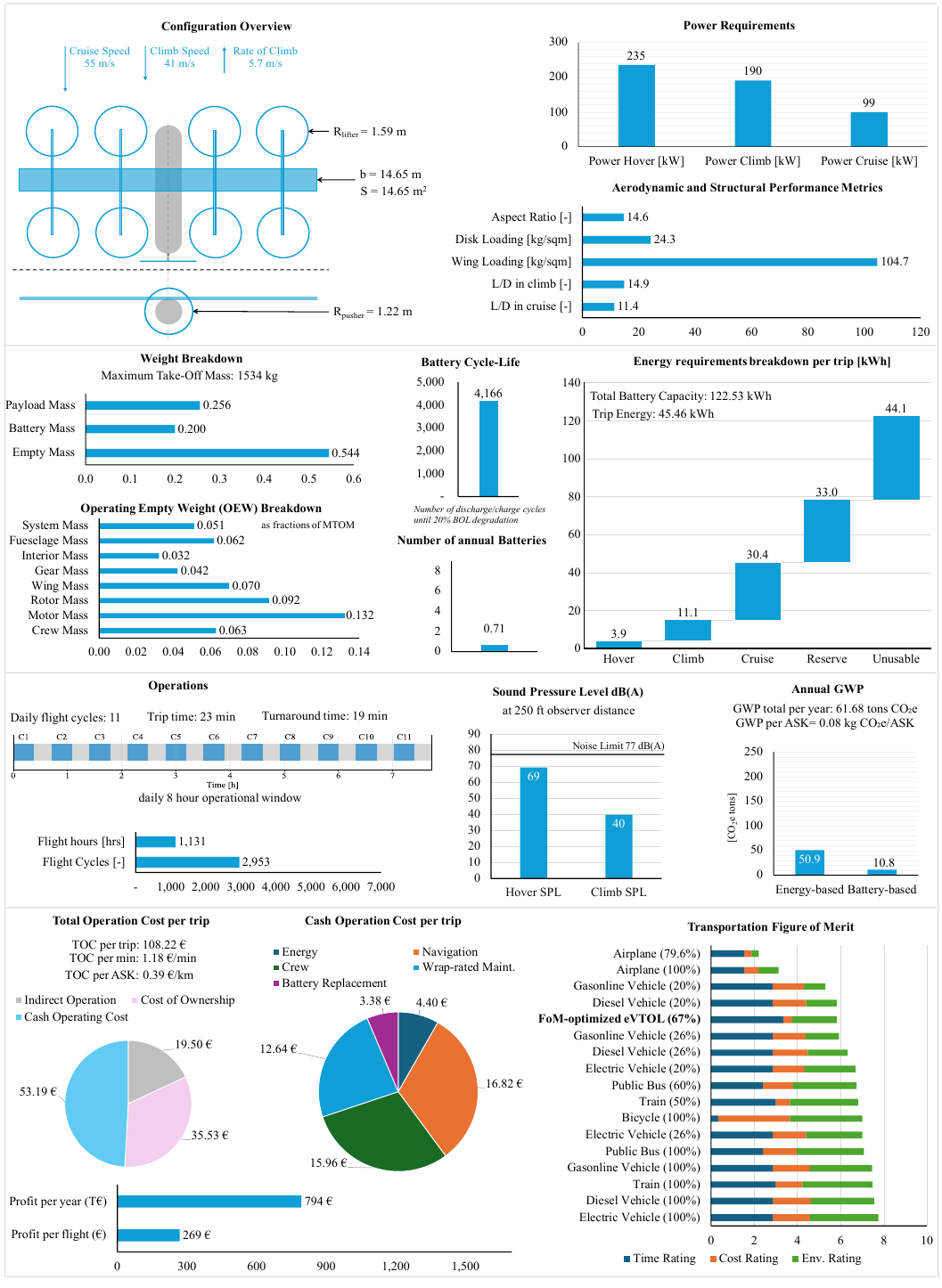}
    \caption[Optimization results of FoM-maximizing conceptual eVTOL design.]{Optimization results of FoM-maximizing conceptual eVTOL design: aircraft design, battery behavior, operating window usage, global warming potential, operating costs and cross-transportation comparison.}
    \label{fig:combined_fom_comparison}
\end{figure}

The FoM-maximizing eVTOL design, illustrated in Figure \ref{fig:combined_fom_comparison}, converged after 4 iterations, 39 function evaluations, and 4 gradient evaluations, completing in 0.57 seconds. The optimizer yields a wing span of 14.65 m with a 1.0 m chord, a cruise propeller radius of 1.22 m, and hover rotors of 1.59 m. The resulting maximum take-off mass is 1,534 kg, with an empty-mass fraction of 0.54 and a battery fraction of 0.20. At a pack-level energy density of 400 Wh/kg, the required capacity of 122.53 kWh corresponds to a battery mass of 306 kg. Cruise is flown at 55 m/s with a lift-to-drag ratio of 11.4, resulting in a flight time of 23 minutes for the 70 km trip at a cruise power of 99 kW. The configuration represents a compromise, with a high aspect ratio wing and moderate rotor diameters, lower disk loading and power demand, while cruise speed remains high enough to ensure acceptable trip duration. The FoM-optimized design is primarily governed by the 15 m vertiport footprint constraint, with the optimizer driving the wingspan to this upper limit to maximize aerodynamic efficiency and lift-to-drag performance, while maintaining the highest margins for noise and weight across the optimized designs (see Table 12 in the supplementary material).

Operational characteristics emphasize balance rather than extremes. Charging at 1.15C leads to a turnaround time of 19 minutes and a block time of 42 minutes per flight. This enables 11 flights per day, translating to 2,952 annual flight cycles, equivalent to 1,130 annual flight hours. Energy use per trip is 45.46 kWh, with a depth-of-discharge of 0.37 and an average discharge C-rate of 0.97 1/h. These values provide a battery life of 4,165 cycles, translating into 0.71 battery replacements per year, or one battery replacement every 1.41 years, respectively. 

The economic profile reflects this balance. Energy costs per trip are €4.39 (4\% of TOC), crew costs €15.96 (15\% of TOC), and navigation €16.82 (16\% of TOC). Maintenance contributes €16.02 (15\% of TOC), while cost of ownership accounts for €35.53 (34\% of TOC), making it the highest cost component. Battery-related costs are minimal at €3.38 per trip (3\% of TOC). The total operating cost is €108.22 per flight, or €0.39 per seat-kilometer. With ticket revenues assumed constant, profit amounts to €268 per flight, yielding €0.793 million annually. 

Environmental performance is shaped by low turnover of batteries and moderate energy demand. Each flight produces 20.89 kg CO$_2$e (54.56 kg CO$_2$e per flight hour), dominated by electricity use, 82\%, and 18\% from battery replacement. On an annual basis, the design emits 61.68 tons CO$_2$e, comparable to 14 gasoline-powered passenger vehicles or 55 electric-powered passenger vehicles driven for one year \cite{EPA2025_GHG_Equivalencies}. The SPL remains within limits at 69 dB(A) during hover. In cross-modal comparison, the design achieves a moderate transportation Figure of Merit, ranking above gasoline-powered and diesel-powered passenger vehicles at 20\% load factor, and right below the same at 26\% load factor. With a trip time of 23.1 minutes, it scores $R_{EVTOL_{time}}=10$ for time. Emissions performance reaches $R_{EVTOL_{CO_{2}e}}=6.22$, while cost is rated at $R_{EVTOL_{cost}}=1.29$. The overall FoM is 5.84, reflecting a configuration that balances energy efficiency, operating cost, and environmental performance.

\paragraph{Key Takeaways:}
\begin{itemize}
    \item Utilizes a high-aspect-ratio wing (14.65 m span) and moderate cruise speed (55 m/s) to balance aerodynamic efficiency with passenger time-value.
    \item Employs a moderate 1.15C charging rate that extends battery life to 4,165 cycles. This results in the low battery-related cost (3\% of TOC) while still maintaining a viable utilization of 11 flights per day.
    \item Achieves the highest transportation FoM across the optimized eVTOL designs by balancing time-saving, emissions, and cost performance, effectively outperforming conventional gasoline and diesel ground vehicles when operated at a low (20\%) passenger load factor.
\end{itemize}

\section{Comparative Assessment of Optimized eVTOL Designs}
\label{ch:discussion}

To assess the modeling capability of the optimization framework, the FoM-optimized design is benchmarked against the multidisciplinary model of Brown and Harris \cite{brown2020}, as its balanced objective (optimizing across time, cost, and emissions simultaneously) makes it the most broadly comparable design. This comparison (Table \ref{tab:validation}) evaluates the framework’s ability to capture scaling laws by reproducing a mission with a 93 km range and comparable payload.

\begin{table}[h]
\centering
\small
\renewcommand{\arraystretch}{1.3}
\begin{tabular}{l|c|cc}
\hline
\textbf{Metric} & \textbf{Reference \cite{brown2020}} & \textbf{Validation Case*} & \textbf{This Study (FoM-opt.)} \\ \hline \hline
Configuration & Lift+Cruise & Lift+Cruise & Lift+Cruise \\
MTOM [kg] & 1,504 & 1,536 (+2.1\%) & 1,534 (+2.0\%) \\
Empty Mass [kg] & 797 & 830 (+4.1\%) & 835 (+4.8\%) \\
Battery Mass [kg] & 348 & 347 (-0.2\%) & 306 (-12.0\%) \\
Payload [kg] & 359 & 359 (0.0\%) & 393 (+9.4\%) \\
Cruise Speed [km/h] & 215 & 197 (-8.1\%) & 200 (-7.0\%) \\
Trip Cost [€] & 288 & 128 (-55.6\%) & 108 (-62.4\%) \\ \hline
\multicolumn{4}{l}{\textit{*Validation Case replicates the 93 km mission and payload parameters from \cite{brown2020}.}}
\end{tabular}
\caption{Validation of the MDO framework and performance of the FoM-optimized design.}
\label{tab:validation}
\end{table}

The framework demonstrates high fidelity in mass estimation, with the MTOM deviating by only 2.1\%. The slightly higher empty mass (+4.1\%) in this study stems from the regression-based mass models detailed in Section \ref{ch:models}, which yield an empty mass fraction of 0.54 compared to the fixed 0.53 used by Brown and Harris \cite{brown2020}. Despite our more conservative battery end-of-life assumption (64\% usable capacity vs. 80\% in \cite{brown2020}), the battery mass remains consistent (-0.2\%). This is partly attributable to the lower cruise speed produced by our model (-8.1\%), which results from computing cruise speed through lift equilibrium rather than prescribing it as an input assumption \cite{brown2020}.

The most pronounced discrepancy is found in the trip cost (-55.6\%), which is primarily driven by evolving economic assumptions in the UAM sector. While Brown and Harris \cite{brown2020} utilized utility-scale storage battery specific costs ($\approx$ €340/kWh), this study adopts an automotive-derived baseline of €115/kWh \cite{Mihara2021}. Although recent literature suggests that aerospace-grade requirements may keep medium-term battery costs closer to €250/kWh \cite{kirste2025a}, our results underscore the high sensitivity of UAM economic projections to these assumptions, particularly if the industry cannot achieve the price levels of the mass-market EV sector. Furthermore, this study assumes a lower Indirect Operating Cost (IOC) factor of 22\%, based on ICAO airline productivity data \cite{icao_airlines_operating_costs}, compared to the 40\% factor in \cite{brown2020}. Finally, our pilot cost modeling reflects a high-utilization UAM structure with an annual salary base ($\approx$ €40/flight hour) rather than traditional business aviation hourly rates ($\approx$ €85/hour \cite{brown2020}). While our resulting costs represent an optimistic scenario, the overall alignment of the physical parameters confirms that the presented model produces reasonable results. To further benchmark the environmental modeling capability of the framework, our results are compared with recent life-cycle assessments in the literature. Fioriti et al. \cite{Fioriti2024ICAS} reported approximately 49 kg CO$_2$e per flight hour for a 1,825 kg MTOM eVTOL aircraft, accounting for maintenance and energy emissions. In comparison, our GWP-optimized (46 kg) and FoM-optimized (55 kg) designs fall within a consistent range. Conversely, the profit-optimized design (151 kg) is a significant outlier. This discrepancy is primarily driven by the high battery-replacement rate in the profit-maximized scenario, exceeding 6.4 battery packs per year compared to approximately 3 in \cite{Fioriti2024ICAS}. Furthermore, unlike the component-weight-fraction approach used in \cite{Fioriti2024ICAS}, our model accounts directly for battery pack life-cycle emissions, capturing the environmental penalty of high-utilization operational profiles.

The analysis of four specialized eVTOL designs, optimized for maximum annual profit, minimum TOC, minimum annual GWP, and maximum FoM, highlights the trade-offs that arise when operational, economic, and environmental priorities are considered. Each configuration demonstrates specific advantages but also exposes limitations that may constrain its broader applicability. To assess conceptual alignment with UAM requirements, the designs are compared with Uber Elevate specifications \cite{uberelevate, uberelevate2023} in Table \ref{tab:ubercomparison}, focusing on key differences in performance and competitiveness. This comparison is intended to evaluate the deviation between early industry projections and the results of this study, providing a preliminary viability assessment. Finally, we discuss how these outcomes map to stakeholder priorities and their operational and regulatory implications.

\begin{table}[h]
\centering
\small
\renewcommand{\arraystretch}{1.5}
\begin{tabular}{l|c|cccc}
\hline
& \textbf{Uber Elevate \cite{uberelevate}} & \textbf{Profit-opt.} & \textbf{TOC-opt.} & \textbf{GWP-opt.} & \textbf{FoM-opt.} \\ \hline \hline
\textbf{Vehicle Design and Performance} & & & & & \\ \hline
Maximum take-off mass [kg] & 1,814 & 1,784 & 1,633 & 1,559 & 1,534 \\ \hline 
Battery cycle life [cycles] & 2,000 & 845 & 2,015 & 5,723 & 4,166 \\ \hline
Battery capacity [kWh] & 140 & 166 & 162 & 127 & 123 \\ \hline
Cruise speed [km/h] & 320 & 259 & 255 & 154 & 200 \\ \hline
Motion efficiency at cruise [km/kWh] & 1.24 & 1.29 & 1.30 & 1.35 & 1.54 \\ \hline
Power required at hover [kW] & 500 & 295 & 297 & 246 & 235 \\ \hline
Power required at cruise [kW] & 120 & 157 & 150 & 89 & 99 \\ \hline
Lift-to-drag ratio [-] & 11 & 11 & 10 & 10 & 11 \\ \hline \hline

\textbf{Economic Competitiveness} & & & & & \\ \hline
Cost per pax-kilometer [€/km] & 0.27 & 0.39 & 0.34 & 0.46 & 0.39 \\ \hline
Cost per pax-minute travelled [€/min] & 1.27 & 1.52 & 1.30 & 1.11 & 1.18 \\ \hline 
Cost of a 70 km pooled VTOL trip per person [€] & 19 & 41 & 35 & 49 & 40 \\ \hline
Vehicle utilization (hours per year) & 2,080 & 1,634 & 1,324 & 1,136 & 1,131 \\ \hline
Load factor (average seats filled) & 67\% & 67\% & 67\% & 67\% & 67\% \\ \hline \hline

\textbf{Operational Costs} & & & & & \\ \hline
Piloting cost as \% of direct operating costs & 36\% & 10\% & 14\% & 19\% & 18\% \\ \hline
Maintenance costs as \% of direct operating costs & 22\% & 37\% & 25\% & 18\% & 18\% \\ \hline
Battery costs as \% of direct operating costs & 2\% & 37\% & 12\% & 2\% & 4\% \\ \hline
Energy costs as \% of direct operating costs & 12\% & 6\% & 7\% & 5\% & 5\% \\ \hline
Ownership costs as \% of direct operating costs & 8\% & 27\% & 32\% & 43\% & 40\% \\ \hline \hline
\end{tabular}
\vspace{6pt}
\caption{Summary of Uber eVTOL UAM requirements and comparison with optimized designs.}
\label{tab:ubercomparison}
\end{table}

Uber's requirements emphasize motion efficiency and operational capability. All optimized designs fall substantially below Uber's target cruise speed of 320 km/h, with values ranging from 259 km/h (profit-optimized) to 154 km/h (GWP-optimized). Despite this, all four designs exceed the mission-level motion efficiency benchmark of 1.24 km/kWh, with values ranging from 1.29 km/kWh (profit-optimized) to 1.54 km/kWh (FoM-optimized). This corresponds to a mission energy demand of 45.5-54.5 kWh for the 70 km sizing mission compared to approximately 56.5 kWh implied by Uber's benchmark. Additionally, battery cycle life varies substantially across the designs relative to Uber's reference value of 2,000 cycles. The profit-optimized design falls 58\% below this reference at 845 cycles, driven by its aggressive 4C charging strategy, representing an 85\% decrease compared to the GWP-optimized design (5,723 cycles) and an 80\% decrease compared to the FoM-optimized design (4,166 cycles). The TOC-optimized design closely aligns with the reference at 2,015 cycles, while the GWP- and FoM-optimized designs substantially exceed it, demonstrating that conservative charging strategies can significantly extend battery service life relative to early UAM planning assumptions. 

The optimized designs show physical characteristics consistent with industry projections, with MTOM values within a range of 2\% for the profit-optimized design and 10\%-15\% for the TOC-, GWP-, and FoM-optimized designs. The GWP- and FoM-optimized designs achieve lower cruise power requirements, albeit with a significant reduction in cruise speed. This reveals a fundamental distinction between the two objectives at the annual operational scale. The GWP-optimized design achieves the lowest energy consumption per flight hour (96 kWh vs. 106 kWh for the FoM-optimized design), reflecting its lower cruise power at 43 m/s. Despite accumulating comparable annual flight hours (1,136 vs. 1,131), the GWP-optimized design completes 21\% fewer flight cycles per year (2,317 vs. 2,953), as each cycle requires a longer block time driven by the conservative 1C charging strategy. However, none of the designs meet the 2,080 flight hours per year, with the profit-optimized design reaching only 1,634 hours, approximately 450 hours below the benchmark. This shortfall in flight hours, based on the slower cruise speed of the optimized eVTOL designs and charging strategies, could impact service availability and require a larger fleet to meet demand, potentially driving up operational costs. Nonetheless, the turnaround intervals of 5 min (Profit-opt.), 10 min (TOC-opt.), and 19 min (FoM-opt.) align with recent processing assumptions of 5–20 minutes \cite{kirste2025a}, indicating the operational feasibility of these mission profiles. The GWP-optimized design, at 24 minutes, exceeds this range as a direct consequence of the 1C charging strategy, highlighting the inherent trade-off between high-frequency utilization and long-term component sustainability. The implementation of battery swapping strategies could potentially decouple charging times from flight schedules, enabling higher utilization without compromising battery cycle life.

Regarding costs per seat, which Uber sets at €19 for a typical trip, the TOC-optimized design achieves the lowest cost among the specialized designs at €35. This suggests that while the optimized designs are technically viable, reaching the aggressive price points envisioned in early industry targets remains a significant economic challenge for the sector. Potential cost savings are particularly apparent in ownership costs, which are tied to aircraft acquisition expenses. These costs are expected to decrease as production scales, potentially reducing overall operating costs in the long term. Additionally, improvements in battery technology, such as higher specific energy or faster charging methods beyond 1C, could further shift the economic balance, making these designs more viable. Overall, all designs fall within reasonable limits and are broadly aligned with preliminary industry targets, suggesting potential real-world viability. The economic feasibility and operational implementation of these designs merit further research to ensure they can meet the demands of urban air mobility. Moreover, while the environmental benefits of the GWP-optimized design are clear, further assessment is needed to evaluate how these designs integrate into the broader sustainability strategies of emerging UAM operators, particularly in terms of long-term environmental impact and scalability. This comparison illustrates that potential future large-scale UAM operators are aiming for a hard balance in operational efficiency, such as high cruise speed at low power requirements combined with high demands in cost efficiency. 

\begin{figure}[ht]
    \centering
    \includegraphics[width=\textwidth]{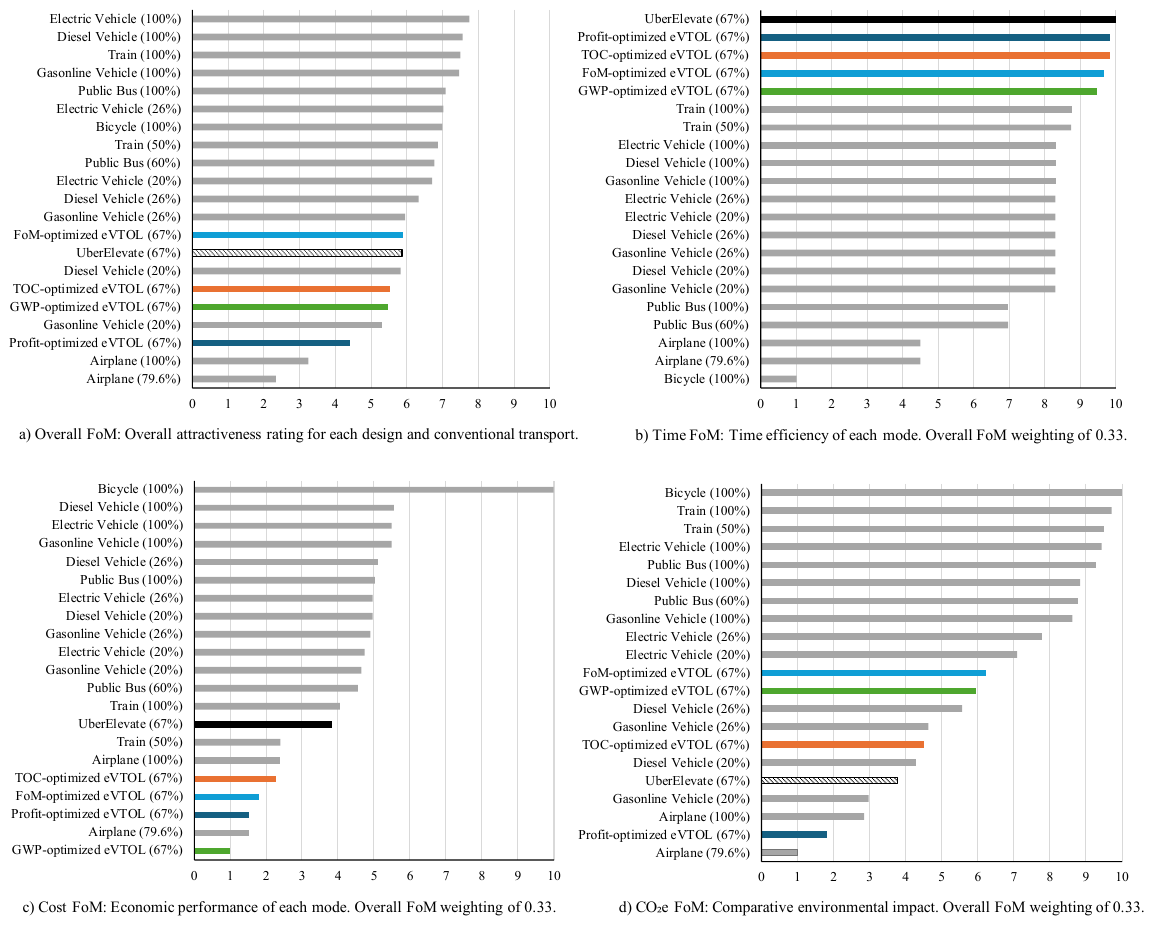} 
    \caption{Comparative assessment of transportation Figure of Merit (FoM) for optimized eVTOL configurations and conventional transport modes as described in Section \ref{ch:fom}. The results, shown for the aggregated FoM (a) and its component dimensions of time (b), cost (c), and environmental impact (d), are benchmarked against industry-recognized UAM requirements derived from Uber Elevate requirements \cite{uberelevate2023}. Source data for this comparison can be found in Table \ref{tab:ubercomparison}. The value presented for Uber Elevate in Figure (c) is a conceptual approximation based on the average of all eVTOL designs, as specific data for this metric was unavailable. It should be noted that this benchmarking serves to position the presented optimized designs against an established industry baseline and should not be taken as a formal evaluation of an Uber Elevate-specific design.}
    \label{fig:figure_foms_compare}
\end{figure}

To understand the overall comparative viability of each design beyond these individual metrics, we introduce the cross-transportation Figure of Merit (FoM). As described in Section \ref{ch:fom}, the FoM addresses this by quantifying the total utility of each transport option, integrating time, cost, and environmental impact into a single rating. As depicted in Figure \ref{fig:figure_foms_compare} (b), all eVTOL designs achieve the highest score for time efficiency, making them the superior choice for time-critical applications \cite{nasaamesresearchcenter2024, zimbelman2023}. In contrast, Figure \ref{fig:figure_foms_compare} (c) reveals their significant comparative disadvantage in cost, where the GWP-optimized design ranks lowest among all options as its slow cruise speed and conservative 1C charging strategy, while effective in reducing battery wear, substantially inflate time-driven costs such as crew and maintenance and simultaneously reduce vehicle utilization, spreading fixed ownership costs across fewer annual flights. However, while even the TOC-optimized design placed in the lower quartile, all optimized designs perform significantly worse than fully-loaded ground-based transport modes like trains or electric passenger vehicles. The Uber Elevate benchmark, with its modest lead over moderate-loaded trains in the cost category, indicates that cost is a critical lever for increasing the overall attractiveness and adoption of eVTOLs beyond niche markets.

Regarding environmental impact, Figure \ref{fig:figure_foms_compare} (d) indicates that optimized eVTOLs perform moderately, except for the profit-maximized design, which scores poorly due to its high annual utilization of batteries and high energy demand. Specifically, prioritizing sustainability in the GWP-optimized design reduces combined energy and battery-related emissions by over 50\% (0.08 kg CO$_2$e/ASK) compared to the profit-maximizing design (0.16 kg CO$_2$e/ASK). Notably, while the GWP-optimized design achieves the lowest annual emissions by minimizing accumulated energy and battery throughput at the annual operational scale, the FoM-optimized design scores marginally higher in the cross-transportation emissions rating, as this metric is evaluated on a per-trip basis where the FoM-optimized design benefits from its higher cruise speed and superior lift-to-drag ratio (11.4 vs. 10.0), resulting in lower energy consumption per flight (45.5 kWh vs. 51.9 kWh) that outweighs its slightly higher battery replacement frequency. The Uber Elevate GWP rating is a conceptual approximation based on the average of the four optimized designs, as specific data were unavailable. The value is not intended as a formal evaluation of the Uber Elevate design. The environmental performance of any eVTOL is heavily dependent on the energy grid source and overall system efficiency. 

Beyond technical metrics, the commercial viability of these designs depends on the passenger's willingness to pay (WTP) for time savings. Recent work by Ison et al. \cite{Ison2024} suggests a general WTP threshold where 62.5\% time savings justify the transition from cars to eVTOLs. The FoM-optimized design achieves a 64\% time saving (41 minutes) compared to a personal vehicle on a 70 km mission, meeting this threshold. However, the attractiveness is considerably lower when compared to rail: while the design still offers a 55\% (28 minutes) time saving over a half-loaded train, this saving entails a cost premium of only 18\% (€40 vs. €34), which falls well within the 71.9\% price tolerance reported by Ison et al. \cite{Ison2024}. Against a fully-loaded train (€17 per trip), however, the cost premium rises to 138\% (€40 vs. €17), far exceeding this threshold and likely limiting eVTOL attractiveness to passengers for whom schedule flexibility and time savings outweigh fare considerations. More broadly, while Ison et al. \cite{Ison2024} report a tolerance for a 71.9\% price increase relative to cars, our results indicate a much steeper cost premium of 274\% compared to a single-occupancy gasoline vehicle (€40 vs. €11), suggesting that the WTP threshold for car users represents a more binding constraint than for rail passengers.

From an environmental perspective, Donateo et al. \cite{Donateo2022} highlight that while eVTOLs can outperform road taxis in high-circuity urban conditions using clean energy, advancing electric road vehicle efficiency may consistently outperform air taxis on a per-km emission basis. This is reflected in our FoM results in Figure \ref{fig:figure_foms_compare} (d), where the GWP- and FoM-optimized designs surpass low-occupancy gasoline and diesel vehicles at 20\% passenger load factor in emissions performance, but are already outperformed by low-occupancy electric vehicles, and fall further behind high-utilization public transit and fully-loaded ground vehicles. This indicates that the environmental competitiveness of eVTOLs is primarily dependent on the comparison against fossil-fuel-powered, low-occupancy ground transport, and that the transition to higher electric vehicle adoption rates will progressively erode this advantage. As ground-based electric vehicles become more efficient and better utilized, this comparative gap is likely to widen further. Future research should utilize the FoM framework to perform extended comparative studies or forecast-based assessments as the UAM industry and ground-based competitors continue to evolve.

The overall FoM in Figure \ref{fig:figure_foms_compare} (a) reflects these trade-offs, showing that an objective, rational stakeholder who equally weights time, cost, and environmental impact would favor ground-based transport like electric cars or trains. This highlights that a straightforward equal weighting, while useful for a conceptual baseline comparison, may not fully capture the complexity of specific stakeholder preferences. For example, while the bicycle achieves the highest cost and emissions score, it is not a practical substitute for a 70 km mission, illustrating that equal weighting across criteria can produce results that are formally optimal but operationally meaningless without minimum service-level constraints. Future work should explore the use of variable weightings and vehicle load factors to reflect different stakeholder segments and incorporate additional metrics such as safety, acoustics, and last-mile accessibility to provide a more holistic, stakeholder-specific view.

Our findings on the high cost and environmental dependency of eVTOLs are consistent with the literature \cite{Hagag2023, Perez2025, Donateo2022}, as described in Section \ref{ch:literature}. However, the presented MDO approach for the design of the eVTOL vehicle, integrating operational, economic, and environmental dimensions at the conceptual design stage, offers a unique systems-level insight into the comparative performance of a transport system based on stakeholder-specific design objectives. The presented cross-transportation FoM offers a generalizable framework to systematically benchmark eVTOLs against the broader transport system. Rather than converging to a single universal optimum, the demonstrated analysis highlights that eVTOL design outcomes are inherently dependent upon the weighting of economic, environmental, and operational objectives, as well as the operational scale at which these are evaluated. This underscores the need for stakeholder-specific pathways in conceptual design, and positions the cross-transportation FoM as a flexible tool for navigating these trade-offs across diverse operational contexts.

\section{Conclusion}
\label{ch:conclusion}

Our multidisciplinary design optimization framework provides a platform for analyzing and optimizing technological, environmental, and economic interdependencies of the eVTOL ecosystem. The analysis of the specialized eVTOL designs provides important insights to address complex trade-offs between these domains, necessary to be considered by UAM stakeholders. Our results provide an in-depth assessment of four specialized eVTOL designs, each optimized for different objectives: maximum profit, minimum total cost of ownership (TOC), minimum global warming potential (GWP), and maximum figure of merit (FoM). The main conclusions from our work and UAM stakeholder strategies and recommendations are as follows:

A profit-optimized eVTOL design achieves substantial operational efficiency gains through short travel times and minimized turnaround and battery recovery times, yet its long-term sustainability is hindered by the environmental impact of current battery technologies, highlighting the need for advancements in battery recycling. In contrast, a TOC-optimized design maximizes cost efficiency but sacrifices operational flexibility and profitability, underscoring the importance of balancing cost minimization with operational performance metrics. A GWP-optimized design, while significantly reducing environmental impact, faces limitations in economic viability due to extended travel times and reduced flight cycles, as energy consumption-optimized airspeed and a battery charging management system tailored to longevity are decisive factors. Finally, a cross-transportational comparative FoM-maximized design provides a balanced design approach based on passenger preferences and comparability with ground transportation alternatives. This approach integrates cost efficiency, operational performance, and sustainability to provide a comprehensive solution for urban air mobility requirements.

Slow charging extends battery life and reduces the frequency of replacement, which reduces long-term costs and improves sustainability, as shown in the scenario for minimum GWP- and maximum FoM-eVTOL design. Despite increased turnaround times and potential losses in operational efficiency, the cost and ecological advantages outweigh the disadvantages. Battery swapping stations allow for quick replacement of discharged batteries at vertiports, reducing downtime and maximizing daily flight cycles. This increases profitability through more frequent flights, but comes with risks such as high investment costs, complex inventory management, and the need for standardized eVTOL designs. The regulatory and logistic challenges of integrating battery swapping and charging stations should be addressed in future work. 

Removing vertiport dimension limitations allows for optimized eVTOL designs with larger wingspans and rotor diameters, improving aerodynamic efficiency and reducing energy consumption. However, flight efficiency-optimized eVTOL designs are reaching the limits of existing helipads. Adaptive wing technologies, such as Airbus' Extra Performance Wing demonstrator \cite{airbus2023extra}, offer a solution by utilizing foldable and movable wings that both improve aerodynamics and ensure infrastructure compatibility. This innovation enables efficient use of existing landing sites, but increases technical complexity and requires careful consideration of certification.

In summary, a strategic focus on battery life extension, recycling processes, optimized vertiport utilization, and adaptive wing designs that combine efficient aerodynamic design with infrastructure compatibility is recommended. The further development of eVTOL technologies should be supported by life cycle analyses, precise noise modeling, and economic operational optimization. Future work should consider additional, critical multidisciplinary aspects like aeroelastic instabilities (important for higher aspect ratio wings) and thermal runaway phenomena (battery safety) to fully capture the design complexity of future operational and regulatory frameworks. While our current framework utilizes physics-based conceptual models suitable for the presented trade-off study, we recognize the potential for further model calibration. Future work could explore the integration of NASA's Design and Analysis of Rotorcraft (NDARC) weight technology factors and subsystem mass models \cite{johnson2025b}, as well as the modular mission performance approaches demonstrated in recent literature \cite{johnson2025a, yipa, nguyen2026}. Furthermore, while forward finite-difference approximations provided a computationally efficient approach for the low-dimensional design space explored in this study, future scaling of the eVTOL design framework would benefit from the implementation of analytic derivatives. Transitioning to algorithmic differentiation frameworks such as JAX \cite{jax2018github} is identified as a key area for future work to improve gradient precision and computational efficiency in higher-fidelity MDO applications \cite{pacini2023}. Integrating these models will further refine the framework's ability to quantify the trade-offs between technical performance and operational sustainability. Beyond this, close co-operation between industry and regulatory authorities is essential to make UAM sustainable and economically viable.

\section{Supplementary Material}
\label{ch:sup}

Supplementary material is available for download here: \href{https://github.com/JohannesJanning/evtol_mdo/blob/main/Supplementary%20Material.pdf}{\textbf{Supplementary Material}}. 

The source code and industry design comparison file is available at: \href{https://github.com/JohannesJanning/evtol_mdo}{\textbf{GitHub}}.

\paragraph{Funding Sources:}
This research did not receive any specific grant from funding agencies in the public, commercial, or not-for-profit sectors.

\newpage

\end{document}